\documentclass[11pt,a4paper]{article}

\usepackage[margin=2.4cm]{geometry}
\usepackage{amsmath,amssymb,mathtools}
\usepackage{booktabs}
\usepackage{array}
\usepackage{graphicx}
\usepackage{float}
\usepackage{xcolor}
\usepackage{cite}
\usepackage{hyperref}

\hypersetup{
  colorlinks=true,
  linkcolor=blue!55!black,
  citecolor=blue!55!black,
  urlcolor=blue!55!black
}

\newcommand{\ii}{\mathrm{i}}
\newcommand{\diag}{\operatorname{diag}}
\newcommand{\ImPart}{\operatorname{Im}}

\DeclareMathOperator{\rank}{rank}

\title{\textbf{
Minimally modified Frampton--Glashow--Yanagida ansatz and its leptogenesis consequences: in light of recent JUNO's result
}}

\author{
Xing-Yu Wang$^{1,2}$,
Zhen-hua Zhao$^{1,2,*}$\\[1ex]
\small $^{1}$School of Physics and Electronic Technology,
Liaoning Normal University, Dalian 116029, China\\
\small $^{2}$Center for Theoretical and Experimental High Energy Physics,\\
\small Liaoning Normal University, Dalian 116029, China\\
\small $^{*}$Corresponding author:
\texttt{zhaozhenhua@lnnu.edu.cn}
}

\date{\today}

\usepackage{setspace}
\begin{document}

\maketitle

\begin{abstract}
Motivated by the recent JUNO indication favoring normal neutrino mass
ordering, we study two realizations of the minimally modified
Frampton--Glashow--Yanagida ansatz (where the neutrino Yukawa matrix assumes two texture zeros) in the minimal type-I
seesaw with two right-handed neutrinos. Both retain one Yukawa
texture zero, supplemented by either a single CP-violating
phase (explicit-CP realization) or a remnant CP symmetry (remnant-CP realization). The explicit-CP realization yields
twelve high-energy Yukawa configurations that reduce to three
low-energy classes, with correlated CP phases and predictions for
neutrinoless double-beta decay. In both realizations, a strongly
suppressed second Yukawa entry can emerge, although the full-matrix
hierarchy remains moderate in the quasi-degenerate regime. We analyze
leptogenesis for hierarchical right-handed neutrinos and for quasi-degenerate ones at the TeV and GeV scales. Successful leptogenesis is obtained across a broad range
of the scanned benchmarks, with increased sensitivity to the initial
right-handed-neutrino abundance at the lowest mass scale. Requiring
successful leptogenesis restricts the allowed Dirac phase values while preserving
the possibility of an approximate second zero. The two realizations
exhibit distinct approximate-zero patterns and leptogenesis
predictions, demonstrating how different restrictions on the
high-energy Yukawa structure affect baryogenesis.
\end{abstract}

\clearpage

\section{Introduction}
\label{sec:introduction}

The observation of neutrino oscillations has established that neutrinos
have nonzero masses and that lepton flavors mix
~\cite{Xing:2020ijf}.
These observations require at least two nonzero light-neutrino masses
and therefore provide clear evidence for physics beyond the Standard
Model. Nevertheless, several fundamental properties of the neutrino
sector remain unresolved, including the absolute neutrino-mass scale,
the mass ordering, the nature of neutrinos, and leptonic CP violation.
The Dirac CP phase is still only partially constrained by oscillation
data~\cite{T2K:2019bcf,NuFIT:6.1}, while possible Majorana phases that can
affect lepton-number-violating observables such as neutrinoless double
beta decay are completely unconstrained~\cite{Dolinski:2019nrj}.

Recent results from the Jiangmen Underground Neutrino Observatory
(JUNO) make the neutrino mass ordering issue particularly timely in this
context. With its first \(59.1\) days of data, JUNO achieved
high-precision measurements of the solar oscillation parameters \(\sin^2\theta_{12}\) and \(\Delta m_{21}^2\)~\cite{JUNO:2025gmd}.
More recently, an analysis based on \(207.2\) live days of
JUNO data, combined with long-baseline accelerator constraints
on \(\Delta m_{31}^2\), favors normal ordering (NO) with a
minimum two-sided significance of \(2.3\sigma\)
over the full range of the Dirac CP phase~\cite{JUNO:2026tap}. Beyond the JUNO analysis discussed above, preferences for NO
have also been reported in T2K long-baseline
data~\cite{T2K:2025yoy} and Super-Kamiokande atmospheric
data with reactor constraints on
$\theta_{13}$~\cite{Super-Kamiokande:2023ahc}.
Such preferences are also found in a combined analysis of
Super-Kamiokande, IceCube-DeepCore, and KM3NeT/ORCA atmospheric
data~\cite{Jin:2026ejo}, as well as in global oscillation fits
including SK atmospheric data or the first JUNO
results~\cite{Esteban:2024eli,Esteban:2026phq}.
A mild preference is also found in $\Lambda$CDM fits combining
cosmic microwave background, DESI baryon acoustic oscillation,
and supernova data~\cite{Escudero:2024uea}. Recent theoretical and phenomenological studies motivated by such a NO preference include
grand unified models and constructions of neutrino masses
and mixing~\cite{Chen:2025afg,Chen:2026mvi,Xing:2026lys},
as well as a specific relation between neutrino mass-squared
differences tested against JUNO data~\cite{Alikhanov:2026bhi}.
Motivated by the experimental and cosmological indications
discussed above, we restrict our discussion to NO.

The type-I seesaw mechanism is one of the most plausible frameworks for
generating neutrino masses
~\cite{Minkowski:1977sc,Yanagida:1979as}.
In the type-I seesaw model, the right-handed-neutrino (RHN) masses
and the neutrino Yukawa couplings enter both the seesaw relation
for the light-neutrino mass matrix and the generation and washout
processes that govern the baryon asymmetry through the leptogenesis
mechanism~\cite{Fukugita:1986hr}. Its minimal realization
introduces two RHNs, which are sufficient to
accommodate the two observed neutrino mass-squared differences and
imply one massless light neutrino~\cite{Xing:2020ald}. Henceforth, we confine our discussion to this highly predictive minimal seesaw scenario. Despite this
economical particle content, the neutrino Yukawa matrix cannot be completely
reconstructed from low-energy neutrino measurements alone. For given
RHN masses, the Casas--Ibarra parametrization of the neutrino Yukawa matrix encodes the undetermined
Yukawa freedom in a single complex parameter (see Sec.~\ref{sec:mmfgy_ci})~\cite{Casas:2001sr}.
Additional restrictions on the neutrino Yukawa matrix are therefore
needed to establish predictive connections between low- and
high-energy observables. 

The texture-zero approach provides a straightforward and transparent way of imposing such restrictions.
A well-known example is the Frampton--Glashow--Yanagida (FGY) ansatz,
which is characterized by two texture zeros in the neutrino Yukawa
matrix for the minimal seesaw model~\cite{Frampton:2002qc}. This two-zero structure tightly
constrains the seesaw parameter space and thereby establishes
predictive relations between the low- and high-energy sectors~\cite{Guo:2006qa}. Restrictions on the high-energy sector are also particularly
relevant for leptogenesis.
However, systematic studies have shown that the FGY models are only compatible with inverted ordering, thereby being inconsistent with NO
~\cite{Harigaya:2012bw,Zhang:2015tea,Barreiros:2018ndn,
Zhao:2021tgi}. Approximate versions of these textures have therefore
also been investigated
~\cite{Rink:2016vvl,Qiu:2024typ}. In view of the recent JUNO result favoring NO, it is natural
to investigate whether the predictive structure of the FGY ansatz can
be preserved through a minimal and structured relaxation of its two texture-zero conditions.

In this work, we study two realizations of the minimally modified FGY
(MMFGY) framework, both of which retain one Yukawa zero but assume
different constraints on their CP structure. In the explicit-CP realization,
CP violation originates from a single CP phase in the
Yukawa sector. In the remnant-CP
realization, the CP structure is implicitly constrained by a remnant CP
symmetry. In other words, among the two texture-zero conditions of the
original FGY model, one zero condition is replaced by a restriction on the CP structure.

For the explicit-CP realization, we derive the resulting low-energy
phase correlations and the effective Majorana mass relevant for
neutrinoless double-beta decay. We further examine the possible emergence of an
approximate second Yukawa zero and determine how this behavior is
affected by the requirement of successful leptogenesis. We also
study  leptogenesis phenomenology over different RHN mass
regimes.
This allows leptogenesis to test
high-energy Yukawa configurations that are indistinguishable from
their low-energy phase correlations alone.

The remnant-CP realization combines one exact Yukawa texture zero
with a remnant CP symmetry. Its established low-energy phase
correlations and neutrinoless-double-beta-decay predictions in
Ref.~\cite{Barreiros:2018bju} are not repeated here. Instead, we
focus on its approximate-two-zero behavior and leptogenesis
phenomenology.

The remainder of this paper is organized as follows. In Sec.~2, we
formulate the explicit-CP MMFGY realization and study its low-energy
predictions, approximate-zero behavior, and leptogenesis phenomenology.
In Sec.~3, we investigate the corresponding high-energy issues
for the remnant-CP MMFGY realization. Finally, our main results are
summarized in Sec.~4.

\section{Explicit-CP MMFGY}
\label{sec:mmfgy}

\subsection{Theoretical framework}
\label{sec:mmfgy_theory}

\subsubsection{Texture}

We consider the minimal type-I seesaw model with two right-handed
neutrino fields \(N_{iR}\), with \(i=1,2\). The relevant Yukawa and
mass terms in the lepton sector are
\begin{equation}
  \mathcal{L}
  \supset
  -\overline{L_{\alpha L}}
   (Y_\ell)_{\alpha\beta}H\ell_{\beta R}
  -\overline{L_{\alpha L}}
   Y_{\alpha i}\widetilde H N_{iR}
  -\frac{1}{2}
   \overline{N_{iR}^{\,c}}
   (M_R)_{ij}N_{jR}
  +\mathrm{h.c.},
  \label{eq:seesaw_lagrangian}
\end{equation}
where \(L_{\alpha L}\) and \(\ell_{\beta R}\) denote the left-handed
lepton doublets and right-handed charged-lepton singlets,
respectively, \(H\) is the Standard Model Higgs doublet, and
\(\widetilde H=\ii\sigma_2H^*\). The matrices \(Y_\ell\), \(Y\), and
\(M_R\) denote the charged-lepton Yukawa matrix, the neutrino Yukawa
matrix, and the RHN Majorana mass matrix, respectively.

We work in the basis where the charged-lepton and RHN mass matrices
are diagonal and real,
\begin{equation}
  M_\ell=\diag(m_e,m_\mu,m_\tau),
  \qquad
  M_R=D_N=\diag(M_1,M_2).
\end{equation}
In this basis, the generic neutrino Yukawa matrix takes the form
\begin{equation}
  Y=
  \begin{pmatrix}
    Y_{e1}&Y_{e2}\\
    Y_{\mu1}&Y_{\mu2}\\
    Y_{\tau1}&Y_{\tau2}
  \end{pmatrix}.
\end{equation}

The light-neutrino mass matrix is given by
\begin{equation}
  M_\nu=-v^2YD_N^{-1}Y^T,
\end{equation}
where \(v=\langle H^0\rangle\) is the Higgs vacuum expectation value.
Since \(Y\) has only two columns, \(\rank(M_\nu)\leq2\), implying
at least one massless light neutrino. Focusing on normal ordering,
we have
\begin{equation}
  m_1=0,\qquad
  m_2=\sqrt{\Delta m_{21}^2},\qquad
  m_3=\sqrt{\Delta m_{31}^2}.
\end{equation}
Here \(m_i\) denote the light-neutrino masses, and
\(\Delta m_{ij}^2\equiv m_i^2-m_j^2\).
The mass matrix is diagonalized according to
\begin{equation}
  U^\dagger M_\nu U^*=D_\nu,
  \qquad
  D_\nu=\diag(0,m_2,m_3).
\end{equation}
The mixing matrix \(U\) is parameterized as
\begin{equation}
  U=
  V_{\rm PDG}
  (\theta_{12},\theta_{13},\theta_{23},\delta)
  \diag(1,e^{i\sigma},1),
\end{equation}
where \(\delta\) and \(\sigma\) denote the Dirac and physical
Majorana phases, respectively.

Having parametrized the low-energy neutrino sector, we now count
the independent parameters of \(Y\). This \(3\times2\) Yukawa
matrix has six complex entries but contains only nine physical
real parameters, because lepton-field rephasings can always remove three
unphysical Yukawa phases. By contrast, the light-neutrino sector
contains seven physical parameters: two nonzero masses, three
mixing angles, and two CP phases. Thus, for given RHN masses,
two real degrees of freedom in \(Y\) remain undetermined by
low-energy data. The Casas--Ibarra parametrization of $Y$ encodes them
in one complex angle (see Sec.~\ref{sec:mmfgy_ci}).
Including the two RHN masses gives eleven physical real parameters
in total.

The original FGY ansatz restricts this parameter space by imposing
two Yukawa texture zeros, corresponding to four real
conditions. The six FGY textures, with the two zeros located in
different flavor rows and different RHN columns, can be written as
\begin{equation}
\begin{aligned}
T_1&=
\begin{pmatrix}
0 & \times\\
\times & 0\\
\times & \times
\end{pmatrix},
&
T_2&=
\begin{pmatrix}
0 & \times\\
\times & \times\\
\times & 0
\end{pmatrix},
&
T_3&=
\begin{pmatrix}
\times & \times\\
0 & \times\\
\times & 0
\end{pmatrix},
\\[2mm]
T_4&=
\begin{pmatrix}
\times & 0\\
0 & \times\\
\times & \times
\end{pmatrix},
&
T_5&=
\begin{pmatrix}
\times & 0\\
\times & \times\\
0 & \times
\end{pmatrix},
&
T_6&=
\begin{pmatrix}
\times & \times\\
\times & 0\\
0 & \times
\end{pmatrix}.
\end{aligned}
\label{eq:two_zero_textures}
\end{equation}
Here \(\times\) denotes a nonzero entry, which may in general be
complex.

In the Casas--Ibarra parametrization, one texture-zero condition
can be used to fix the complex orthogonal angle up to discrete
branch choices, while the other subsequently yields a complex
relation among the low-energy neutrino parameters. The low-energy
CP phases are therefore no longer independent once the measured
values of mixing angles and mass-squared differences are input.
At the same time, the reduced high-energy freedom allows the CP
violation relevant for leptogenesis to be related to the low-energy
leptonic CP phases. However, with current oscillation constraints,
the FGY textures are incompatible with normal
ordering~\cite{Zhang:2015tea,Zhao:2021tgi}.

Here we seek a minimal relaxation of the FGY texture
by retaining only one Yukawa zero. The resulting one-zero Yukawa
matrix contains two independent physical phases, which can be
represented by the relative phases between the two entries in each
of the two flavor rows without a texture zero. To examine their
respective roles, we study the two possibilities in which each
relative phase acts in turn as the sole source of CP violation. Similar mechanisms for generating such a phase from spontaneous
CP breaking, and for transmitting it selectively to a particular
Yukawa entry, have been discussed in
Refs.~\cite{King:2016yvg,Tanimoto:2024nwm}.

After suitable lepton-field rephasings, the six configurations
with the texture zero in the first RHN column can be written as
\begin{equation}
\begin{aligned}
\mathcal H_{1;e\mu}:&
\quad
Y=
\begin{pmatrix}
0 & \times\\
\times & \ast\\
\times & \times
\end{pmatrix};
\hspace{1cm}
&
\mathcal H_{1;e\tau}:&
\quad
Y=
\begin{pmatrix}
0 & \times\\
\times & \times\\
\times & \ast
\end{pmatrix};
\\[3mm]
\mathcal H_{1;\mu e}:&
\quad
Y=
\begin{pmatrix}
\times & \ast\\
0 & \times\\
\times & \times
\end{pmatrix};
\hspace{1cm}
&
\mathcal H_{1;\mu\tau}:&
\quad
Y=
\begin{pmatrix}
\times & \times\\
0 & \times\\
\times & \ast
\end{pmatrix};
\\[3mm]
\mathcal H_{1;\tau e}:&
\quad
Y=
\begin{pmatrix}
\times & \ast\\
\times & \times\\
0 & \times
\end{pmatrix};
\hspace{1cm}
&
\mathcal H_{1;\tau\mu}:&
\quad
Y=
\begin{pmatrix}
\times & \times\\
\times & \ast\\
0 & \times
\end{pmatrix}.
\end{aligned}
\label{eq:mmfgy_H1_explicit}
\end{equation}
In Eq.~\eqref{eq:mmfgy_H1_explicit}, each \(\times\) denotes an
independent nonzero real entry of either sign, while \(\ast\)
denotes the only potentially complex nonzero entry, whose phase
is the sole source of CP violation.
In \(\mathcal H_{i;\alpha\beta}\), \(i\) and \(\alpha\) specify
the column and flavor row of the texture zero, respectively, while
\(\beta\) labels the row containing the \(\ast\) entry. The phase can equivalently be placed in either nonzero entry of the \(\beta\) row, since the two choices are related by a lepton-field rephasing.
The remaining flavor row, distinct from both \(\alpha\) and
\(\beta\), is denoted by \(\gamma\) and has two real entries.
Thus, for each fixed zero position, the two configurations
displayed side by side correspond to the two choices of the
sole CP source.
The six \(\mathcal H_{2;\alpha\beta}\) configurations are obtained
by interchanging the two columns of the displayed patterns.

In terms of real degrees of freedom, the textures of \(Y\)
considered here reduce the number of independent real parameters
from eleven to eight. As will become explicit below, the
texture-zero condition can be used to fix the two real degrees of freedom contained in the complex
Casas--Ibarra angle, while the single-source CP restriction yields
a correlation between the two low-energy CP phases.

There are two possible RHN-column placements of the texture zero,
three choices of the texture zero flavor \(\alpha\), and two choices of
the CP-violating source phase. The explicit-CP realization therefore
contains
\begin{equation}
  2\times3\times2=12
\end{equation}
distinct high-energy Yukawa configurations. Their low-energy
equivalences will be established below.

\subsubsection{Casas--Ibarra analysis}
\label{sec:mmfgy_ci}

For the minimal seesaw model, the Casas--Ibarra parametrization~\cite{Casas:2001sr} expresses the
neutrino Yukawa matrix as
\begin{equation}
  Y
  =
  \frac{i}{v}
  U D_{\nu}^{1/2}RD_{N}^{1/2},
  \label{eq:CI}
\end{equation}
where \(R\) is a complex orthogonal matrix satisfying
\begin{equation}
  R^{T}R=\mathbf{1}_{2\times2}.
\end{equation}

For normal ordering with \(m_1=0\), the matrix \(R\) can be
parametrized as
\begin{equation}
  R_{\rm NO}
  =
  \begin{pmatrix}
    0 & 0 \\
    \cos z & -\sin z \\
    \xi\sin z & \xi\cos z
  \end{pmatrix},
  \qquad
  \xi=\pm1 ,
\end{equation}
where \(z=x+iy\) is a complex angle, while \(\xi\) is a discrete sign.
Substituting \(R_{\rm NO}\) into Eq.~\eqref{eq:CI} yields the two elements in
the \(\alpha\)-th row of the neutrino Yukawa matrix as
\begin{align}
  Y_{\alpha1}
  &=
  \frac{i\sqrt{M_1}}{v}
  \left(
    \sqrt{m_2}U_{\alpha2}\cos z
    +
    \xi\sqrt{m_3}U_{\alpha3}\sin z
  \right),
  \label{eq:Yalpha1}
  \\
  Y_{\alpha2}
  &=
  \frac{i\sqrt{M_2}}{v}
  \left(
    -\sqrt{m_2}U_{\alpha2}\sin z
    +
    \xi\sqrt{m_3}U_{\alpha3}\cos z
  \right).
  \label{eq:Yalpha2}
\end{align}

Since the texture zero may occur in either RHN column, there are two
possible placements. For \(Y_{\alpha1}=0\),
Eq.~\eqref{eq:Yalpha1} gives
\begin{equation}
  \tan z
  =
  -\frac{\sqrt{m_2}U_{\alpha2}}
  {\xi\sqrt{m_3}U_{\alpha3}},
  \label{eq:tan_z_zero1}
\end{equation}
whereas \(Y_{\alpha2}=0\) gives
\begin{equation}
  \tan z
  =
  \frac{\xi\sqrt{m_3}U_{\alpha3}}
  {\sqrt{m_2}U_{\alpha2}} .
  \label{eq:tan_z_zero2}
\end{equation}
Thus, in either case, the texture-zero condition removes the freedom associated with the complex Casas--Ibarra parameter \(z\), up
to discrete branch choices.

We now derive the low-energy phase relation implied by the
single-phase textures. In $\mathcal H_{i;\alpha\beta}$, the
relative phase in the $\beta$ row is the sole source of CP
violation, while the two entries in the remaining $\gamma$
row are real as given by
Eq.~\eqref{eq:mmfgy_H1_explicit}. Hence,
$Y_{\gamma1}Y_{\gamma2}^{*}$ is real, independently of
lepton-field rephasings.

For the placement $Y_{\alpha1}=0$, substituting
Eq.~\eqref{eq:tan_z_zero1} into the Yukawa expressions gives
\begin{align}
  Y_{\gamma1}Y_{\gamma2}^{*}
  =
  \mathcal C_{\alpha}
  &\left(
    U_{\gamma2}U_{\alpha3}
    -
    U_{\gamma3}U_{\alpha2}
  \right)
  \nonumber\\
  &\times
  \left(
    m_2U_{\gamma2}^{*}U_{\alpha2}^{*}
    +
    m_3U_{\gamma3}^{*}U_{\alpha3}^{*}
  \right),
  \label{eq:Ygamma_product}
\end{align}
where $\mathcal C_{\alpha}$ is a nonzero real factor.
Since \(Y_{\gamma1}Y_{\gamma2}^{*}\) is real, the right-hand side of
Eq.~\eqref{eq:Ygamma_product} must also be real.
Removing the nonzero real factor $\mathcal C_\alpha$,
we obtain the low-energy phase relation
\begin{equation}
\begin{aligned}
  F_{\alpha\beta}
  \equiv \ImPart\Big[
  &\left(
    U_{\gamma2}U_{\alpha3}
    -U_{\gamma3}U_{\alpha2}
  \right)
  \\
  &\times
  \left(
    m_2U_{\gamma2}^{*}U_{\alpha2}^{*}
    +m_3U_{\gamma3}^{*}U_{\alpha3}^{*}
  \right)
  \Big]
  =0 .
\end{aligned}
\label{eq:F_definition}
\end{equation}
Here the indices $(\alpha,\beta)$ follow the labels of
$\mathcal H_{i;\alpha\beta}$, with $\gamma$ denoting the
remaining flavor.

Repeating the same procedure for the placement $Y_{\alpha2}=0$
leads to the same condition $F_{\alpha\beta}=0$. The position of the texture zero in the two RHN
columns therefore does not affect the resulting low-energy
constraint.

The function $F_{\alpha\beta}$ depends only on the light-neutrino
masses, mixing angles, and CP phases. In particular, the RHN masses and
the Casas--Ibarra parameter do not enter the constraint. We may
therefore write
\begin{equation}
  F_{\alpha\beta}
  \left(
    \theta_{ij},m_i,\delta,\sigma
  \right)
  =0 .
\end{equation}
For given values of neutrino masses and mixing angles, this reduces to a relation
between the two physical CP phases,
\begin{equation}
  F_{\alpha\beta}(\delta,\sigma)=0 .
  \label{eq:phase_relation_general}
\end{equation}


\subsubsection{Low-energy classification}
\label{sec:low_energy_classification}

The twelve distinct high-energy Yukawa configurations do not lead to
twelve independent constraints on the low-energy neutrino parameters.
Two high-energy configurations are equivalent at low energies if they lead to the
same constraint on the low-energy parameters.

The first reduction follows from the position of the texture zero in the
RHN columns. As shown in the previous subsection, for a fixed ordered
flavor assignment \((\alpha,\beta)\), placing the texture zero in the
first or second RHN column leads to the same low-energy constraint.
Thus, the configurations with \(Y_{\alpha1}=0\) and
\(Y_{\alpha2}=0\) belong to the same low-energy class, reducing the
twelve high-energy configurations to six independent ordered flavor
assignments.

A further reduction follows from the flavor structure of the
constraint function defined in Eq.~\eqref{eq:F_definition}. Under the
interchange \(\alpha\leftrightarrow\gamma\), with the CP-sourcing
flavor \(\beta\) unchanged, its first factor changes sign,
\begin{equation}
  U_{\alpha2}U_{\gamma3}
  -
  U_{\alpha3}U_{\gamma2}
  =
  -
  \left(
    U_{\gamma2}U_{\alpha3}
    -
    U_{\gamma3}U_{\alpha2}
  \right),
\end{equation}
whereas the second factor is symmetric. Consequently,
\begin{equation}
  F_{\gamma\beta}
  =
  -F_{\alpha\beta},
\end{equation}
and hence
\begin{equation}
  F_{\alpha\beta}=0
  \qquad\Longleftrightarrow\qquad
  F_{\gamma\beta}=0 .
  \label{eq:flavor_equivalence}
\end{equation}

The six ordered flavor assignments therefore form three equivalent
pairs,
\begin{equation}
  (e,\tau)\sim(\mu,\tau),
  \qquad
  (e,\mu)\sim(\tau,\mu),
  \qquad
  (\mu,e)\sim(\tau,e).
\end{equation}
We denote the corresponding three low-energy equivalence classes by
the representative ordered flavor pairs
\begin{equation}
  (e,\tau),
  \qquad
  (e,\mu),
  \qquad
  (\mu,e).
\end{equation}
In each label, the first flavor specifies the row containing the
texture zero, while the second specifies the row whose relative phase
is the sole source of CP violation. The chosen pair serves as a
representative label for the corresponding equivalence class.
For example, the class \((e,\tau)\) contains the configurations with
ordered flavor assignments \((e,\tau)\) and \((\mu,\tau)\), for either
choice of the RHN column containing the texture zero.

The three independent low-energy constraints can therefore be
represented by
\begin{equation}
  F_{e\tau}(\delta,\sigma)=0,
  \qquad
  F_{e\mu}(\delta,\sigma)=0,
  \qquad
  F_{\mu e}(\delta,\sigma)=0 .
  \label{eq:three_low_energy_classes}
\end{equation}
Thus, the twelve distinct high-energy Yukawa configurations reduce to
three equivalence classes for observables determined entirely by the
low-energy neutrino parameters. This low-energy equivalence does not,
however, imply identical Yukawa matrices. The distinct high-energy
configurations must therefore still be treated separately for
quantities that depend explicitly on the Yukawa structure, in
particular leptogenesis.

\subsection{Phenomenological results}
\label{sec:mmfgy_pheno}

\subsubsection{Low-energy predictions}
\label{sec:mmfgy_low}

\begin{figure}[htbp]
  \centering
  \includegraphics[width=0.70\textwidth]{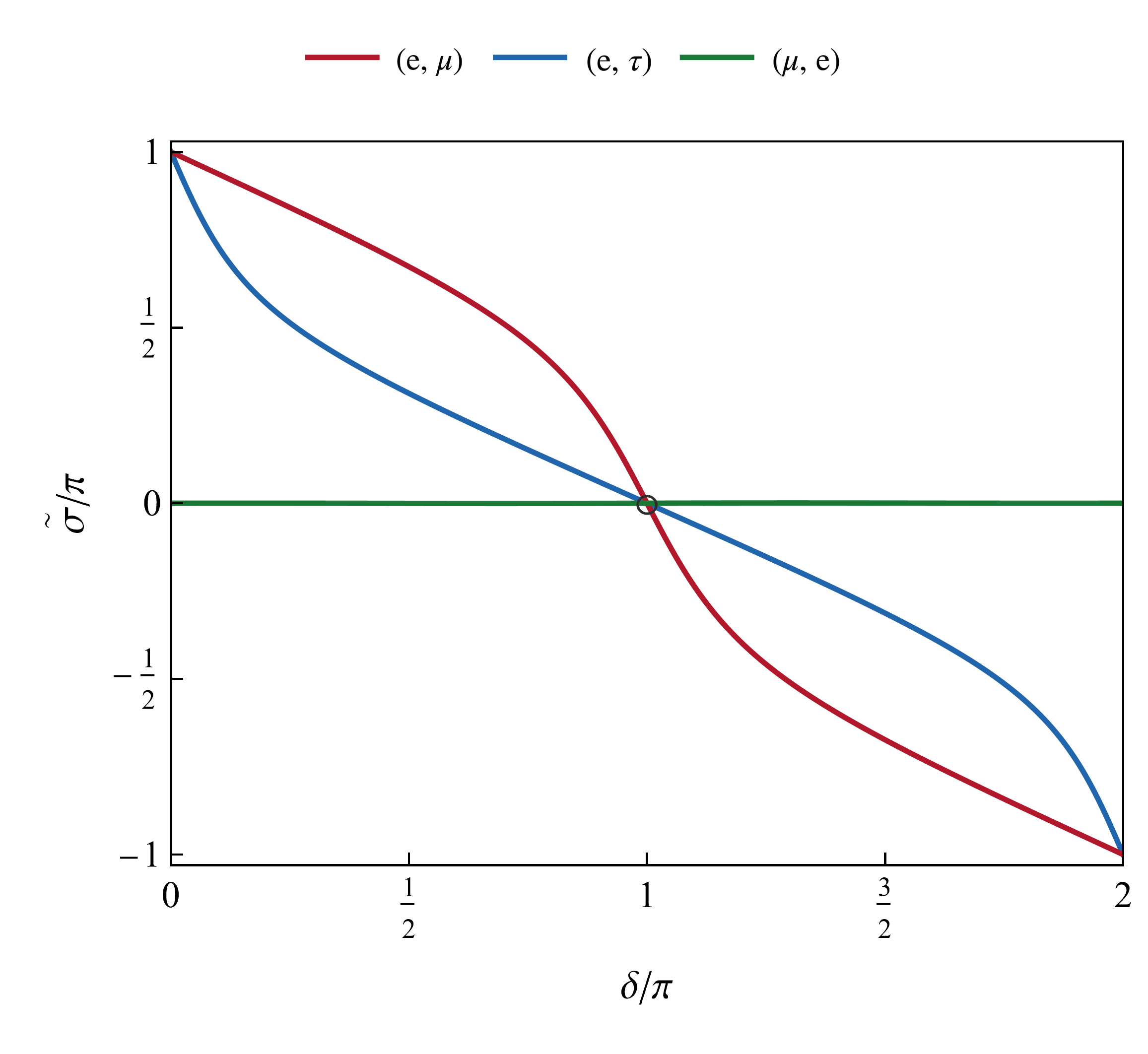}
  \caption{
Correlations between the Dirac phase \(\delta\) and the continuous
representative \(\widetilde{\sigma}\) of the Majorana phase for the
three independent low-energy classes of the minimally modified FGY
ansatz, evaluated using the NuFIT 6.1 best-fit inputs.
}
  \label{fig:phase}
\end{figure}

Having established the three independent low-energy classes, we now
examine their predictions for the two physical CP phases. For given values of
neutrino masses and mixing angles, each constraint in
Eq.~\eqref{eq:three_low_energy_classes} determines the Majorana phase
\(\sigma\) as a function of the Dirac phase \(\delta\). Since
\(\sigma\) is periodic modulo \(\pi\), for visualization we unwrap the
phase and denote its continuous representative by
\(\widetilde{\sigma}\), with the branch chosen such that
\(\widetilde{\sigma}=0\) at the common solution \(\delta=\pi\).
The resulting correlations are shown in Fig.~\ref{fig:phase}.
In the numerical calculations, we use the NuFIT 6.1 global analysis results [3] as the oscillation
input. Unless stated otherwise, the NuFIT 6.1 best-fit values are used.

All three classes pass through the common CP-conserving solution
\begin{equation}
  \delta=\pi,
  \qquad
  \sigma=0 .
\end{equation}
Away from this point, the \((e,\tau)\) and \((e,\mu)\) classes require
\(\sigma\) to vary with \(\delta\), but with different slopes. This
difference arises because the muon- and tau-flavor components enter
the corresponding conditions \(F_{e\tau}=0\) and \(F_{e\mu}=0\) with
different mixing-angle weights. By contrast, the $(\mu,e)$
trajectory remains close to
$\sigma=0$ throughout the scanned range of $\delta$.
This is because the $\delta$-dependent terms in the corresponding constraint
are suppressed by the small mixing angle $\theta_{13}$,
with the leading contribution further reduced by
$\theta_{23}$ being close to $45^\circ$.
The three low-energy classes therefore predict distinct
relations between the two physical CP phases.

These phase relations also determine the effective Majorana mass
relevant for neutrinoless double-beta decay,
\begin{equation}
  \langle m_{ee}\rangle
  =
  \left|
    \sum_{i=1}^{3} m_i U_{ei}^{2}
  \right|.
  \label{eq:mee}
\end{equation}
For normal ordering with \(m_1=0\), this reduces to
\begin{equation}
  \langle m_{ee}\rangle
  =
  \left|
    m_2s_{12}^{2}c_{13}^{2}e^{2i\sigma}
    +
    m_3s_{13}^{2}e^{-2i\delta}
  \right|.
  \label{eq:mee_NO}
\end{equation}
Along each allowed phase trajectory, substituting the corresponding
$\sigma(\delta)$ into Eq.~\eqref{eq:mee_NO} gives a one-parameter
prediction for $\langle m_{ee}\rangle$. The relative phase of the two
contributions is $2[\sigma(\delta)+\delta]$, so the three phase
trajectories lead to different amounts of addition or cancellation
between them.

\begin{figure}[htbp]
  \centering
  \includegraphics[width=0.99\textwidth]{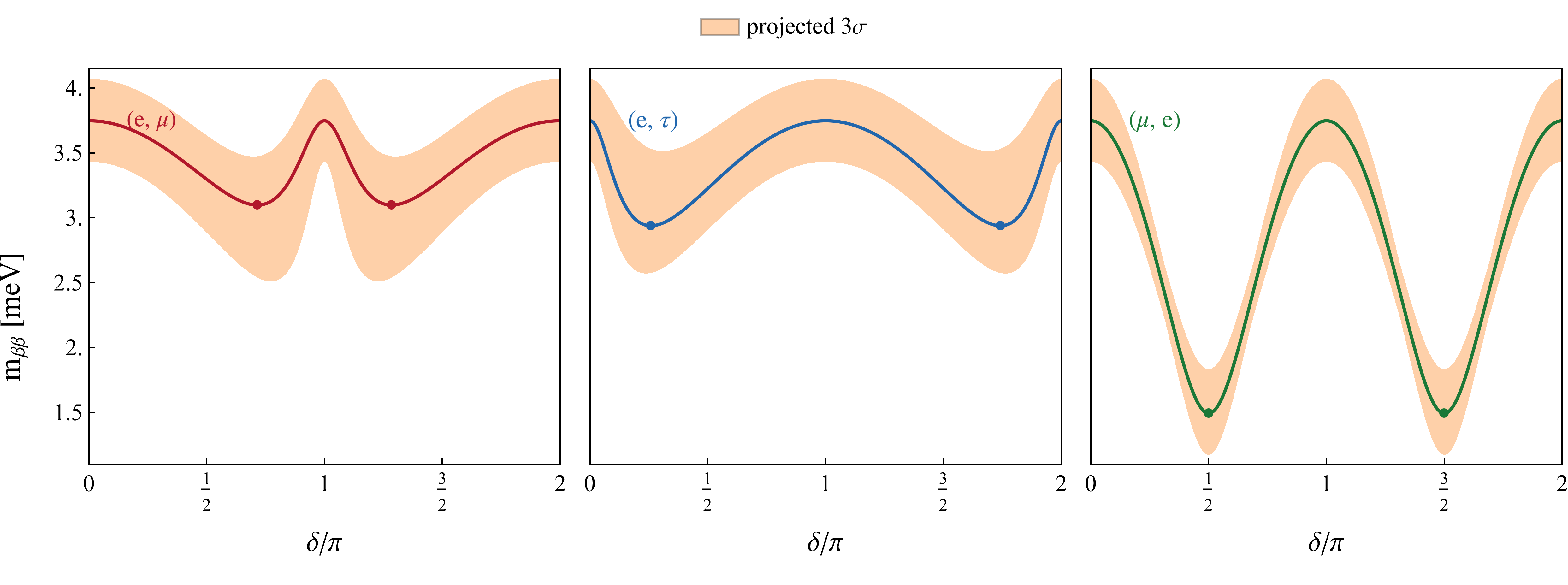}
  \caption{
    Effective Majorana mass \(\langle m_{ee}\rangle\) as a function of
the Dirac phase \(\delta\) for the three independent low-energy
classes.
The solid curves correspond to the NuFIT 6.1 best-fit inputs. The shaded bands show the ranges obtained from the endpoint
    combinations of the one-dimensional \(3\sigma\) intervals of the
    three mixing angles and two mass-squared differences, with the
    phase relation re-solved for each input set.
  }
  \label{fig:mbb}
\end{figure}

As shown in Fig.~\ref{fig:mbb}, all three classes predict
\(\langle m_{ee}\rangle\) at the \(10^{-3}\,{\rm eV}\) scale expected
for normal ordering with \(m_1=0\), but with different dependences on
\(\delta\). At the common solution
\(\delta=\pi\) and \(\sigma=0\), both phase factors in
Eq.~\eqref{eq:mee_NO} equal unity. The two contributions are therefore
in phase and add in magnitude, causing the three best-fit curves to
meet at the same maximum. Away from this point, each class predicts a
different relative phase between the two contributions and hence a
different curve shape.

The widths of the shaded bands have two origins. Variations of the
oscillation parameters in their \(3\sigma\) intervals directly change the magnitudes of the two
terms in Eq.~\eqref{eq:mee_NO}. They also change the value of $\sigma$ determined by the
low-energy phase relation in
Eq.~\eqref{eq:three_low_energy_classes}, thereby shifting
the relative phase of the two terms. In regions where the two
contributions partially cancel, these small input variations can
produce a comparatively larger change in
\(\langle m_{ee}\rangle\).

\subsubsection{Approximate zeros}
\label{sec:mmfgy_approx_zero}

Having obtained the viable normal-ordering solutions, we now quantify
how closely their high-energy Yukawa structures approach the original
two-zero FGY textures. Although the explicit-CP realization retains
only one Yukawa zero in each single-phase configuration, a
candidate second-zero entry may be strongly suppressed compared to other nonzero entries. We therefore
examine whether an approximate second zero can emerge without being
imposed as a model condition.

For a configuration \(\mathcal H_{i;\alpha\beta}\) with
\(Y_{\alpha i}=0\), a candidate second zero of an FGY-type texture must
lie in the other RHN column \(j\neq i\) and in a different flavor row
\(\rho\neq\alpha\). We compare its magnitude only with the other
entries in the same column, since the two RHN columns are associated with
different overall mass factors, \(\sqrt{M_1}\) and
\(\sqrt{M_2}\). This comparison isolates the
relative suppression of the candidate entry within the couplings of
the same RHN. We quantify this suppression by
\begin{equation}
  \epsilon_{\rho j}(\delta)
  \equiv
  \max_{\kappa\in\mathcal S_j(\rho)}
  \frac{|Y_{\rho j}(\delta)|}
       {|Y_{\kappa j}(\delta)|}
  =
  \frac{|Y_{\rho j}(\delta)|}
       {\displaystyle
        \min_{\kappa\in\mathcal S_j(\rho)}
        |Y_{\kappa j}(\delta)|},
  \label{eq:second_zero_ratio}
\end{equation}
where
\begin{equation}
  \mathcal S_j(\rho)
  =
  \left\{
    \kappa\neq\rho
    \,\middle|\,
    Y_{\kappa j}\neq0
  \right\}.
  \label{eq:second_zero_reference_set}
\end{equation}
Thus, \(\epsilon_{\rho j}<1\) means that the candidate is the smallest
nonzero entry in column \(j\), while
\(\epsilon_{\rho j}\ll1\) signals an approximate second zero. Since all entries in
Eq.~\eqref{eq:second_zero_ratio} belong to the same RHN column, their
common factor \(\sqrt{M_j}\) cancels, making the suppression measure
independent of the RHN masses.

\begin{figure}[htbp]
  \centering
  \includegraphics[width=1.00\textwidth]{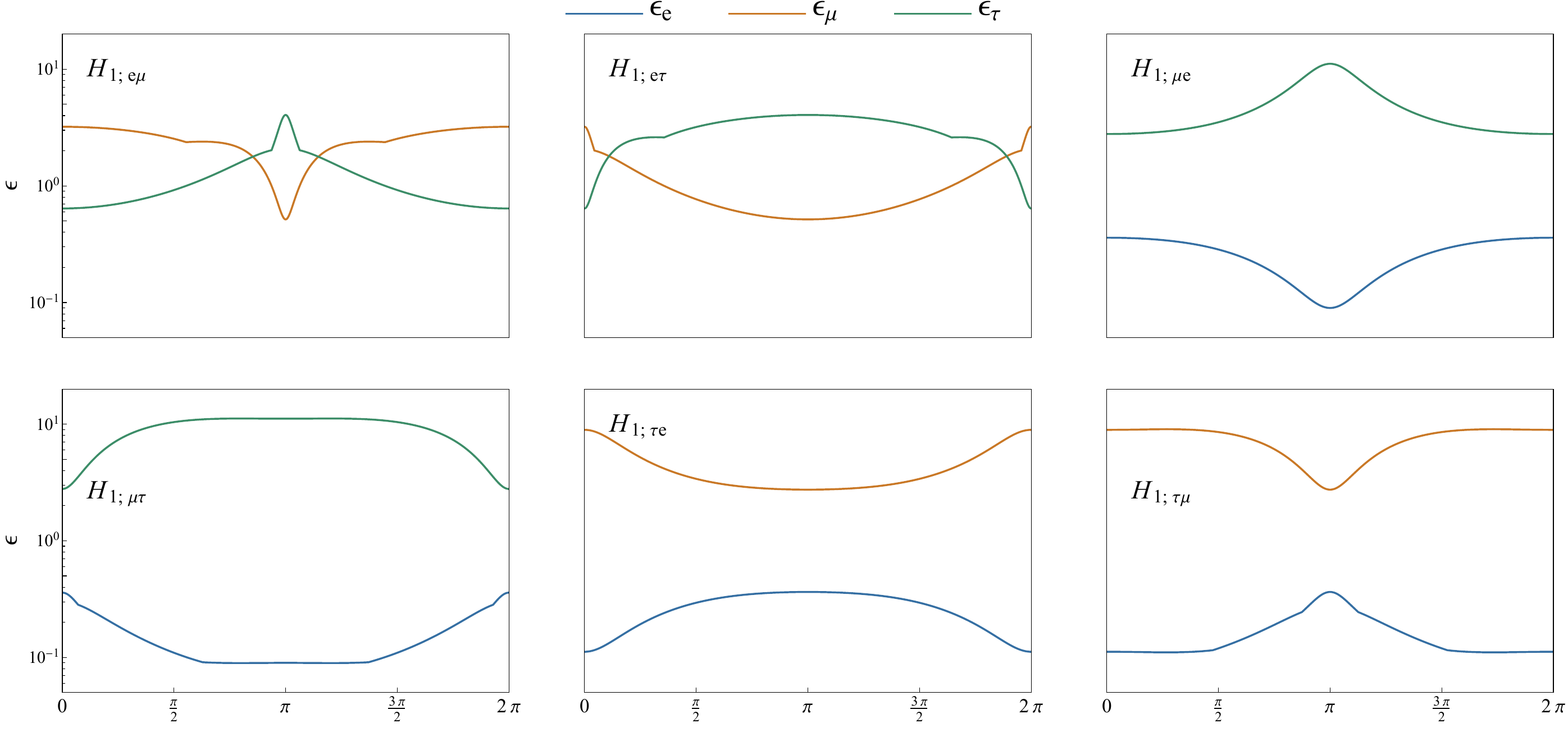}
  \caption{
    Relative suppression \(\epsilon_{\rho j}\) of the candidate
    second-zero entries for the six independent
    \(\mathcal H_{1;\alpha\beta}\) configurations, evaluated along the
    allowed phase correlations using the NuFIT 6.1 best-fit inputs.
    Both possible candidate flavor rows are shown for each
    configuration. The suppression factors are displayed without
    clipping on a logarithmic scale. Values
    \(\epsilon_{\rho j}>1\) indicate that the specified entry is not
    the smallest nonzero element in its RHN column. The corresponding
    \(\mathcal H_{2;\alpha\beta}\) configurations follow by
    interchanging the two RHN columns.
  }
  \label{fig:approx_two_zero}
\end{figure}

The two-zero FGY textures in Eq.~\eqref{eq:two_zero_textures}
form three pairs according to the two flavor rows containing the
zeros. Specifically, \(T_1\) and \(T_4\) contain one zero in the
\(e\) row and one in the \(\mu\) row, \(T_2\) and \(T_5\) contain
zeros in the \(e\) and \(\tau\) rows, and \(T_3\) and \(T_6\)
contain zeros in the \(\mu\) and \(\tau\) rows. Within each pair,
the two textures differ only by interchanging the RHN columns.
Matching the remaining row to the CP-sourcing flavor of each MMFGY
class gives
\begin{equation}
  (e,\tau)\leftrightarrow\{T_1,T_4\},
  \qquad
  (e,\mu)\leftrightarrow\{T_2,T_5\},
  \qquad
  (\mu,e)\leftrightarrow\{T_3,T_6\}.
  \label{eq:fgy_texture_mapping}
\end{equation}
This correspondence applies when the candidate second zero lies
outside the CP-sourcing row; it does not imply that the corresponding
high-energy Yukawa structures are equivalent.

For the \(\mathcal H_{1;\alpha\beta}\) configurations shown in
Fig.~\ref{fig:approx_two_zero}, the closest normal-ordering
solutions occur in the \(T_4\) and \(T_5\) directions, where the minimum suppression factors are approximately
\(0.09\) at \(\delta\simeq143^\circ,\,217^\circ\) and
\(0.11\) at \(\delta\simeq48^\circ,\,312^\circ\), respectively. In these cases,
the candidate entry is at least suppressed by roughly one order of magnitude
relative to other nonzero entries in the same RHN column.
By comparison, the \(T_1\) and \(T_2\) directions reach only about
\(0.52\) and \(0.64\), while the candidates associated with
\(T_3\) and \(T_6\) remain above unity throughout the phase scan
and are therefore never the smallest entries in their columns.
Thus, among these \(\mathcal H_{1;\alpha\beta}\) configurations,
only the \(T_4\) and \(T_5\) directions develop a pronounced
additional Yukawa suppression. The corresponding
\(\mathcal H_{2;\alpha\beta}\) configurations approach
\(T_1\) and \(T_2\), respectively, under interchange of
the RHN columns.

As a complementary full-matrix test, we compare the smallest and
second-smallest magnitudes among all five nonzero Yukawa entries.
This makes sense in the quasi-degenerate regime, where
the mass normalizations of the two RHN columns are approximately
equal and thus the comparison results are approximately independent of the ratio between \(M_1\) and \(M_2\). We define the ratio as
\(r_{\rm global}=y_{(1)}/y_{(2)}\).
Minimizing along the allowed phase trajectories gives
\(r_{\rm global}^{\min}\simeq0.20\), attained at
\(\delta=\pi\) for the \(\mathcal H_{i;\mu e}\) and
\(\mathcal H_{i;\mu\tau}\) configurations with \(i=1,2\).
This corresponds to a strongest full-matrix hierarchy of
approximately \(1:5\).


\subsubsection{Leptogenesis}
\label{sec:leptogenesis}

The constraints derived above make the explicit-CP realization
predictive for leptogenesis. For each distinct Yukawa configuration,
the complete Yukawa matrix is determined once the low-energy inputs,
the Dirac phase, and the RHN masses are specified. The baryon
asymmetry of the Universe therefore provides a high-energy test of the model,
allowing us to examine whether the observed value can be reproduced
and how this requirement constrains the Dirac phase and the RHN mass
parameters.

The relevant leptogenesis dynamics depends strongly on both the absolute scale and the degeneracy degree of RHN masses. For hierarchical RHNs at
high mass scales, standard thermal leptogenesis generates a lepton
asymmetry through out-of-equilibrium RHN decays, with the CP asymmetry
arising from the interference between tree-level amplitudes and their
loop corrections~\cite{Fukugita:1986hr}. For quasi-degenerate RHNs,
the decay asymmetry can be resonantly enhanced when the mass splitting
is comparable to the decay widths, allowing successful resonant
leptogenesis at substantially lower mass
scales~\cite{Pilaftsis:2003gt,Dev:2017wwc}. The dynamics changes qualitatively for RHNs with masses below the electroweak scale \(\sim100~\mathrm{GeV}\): leptogenesis is realized during their out-of-equilibrium production processes rather than decay processes. During this production stage, coherent RHN oscillations can generate lepton asymmetries, a mechanism known as Akhmedov--Rubakov--Smirnov (ARS)
leptogenesis, or leptogenesis via oscillations~\cite{Akhmedov:1998qx}. In all these cases, the asymmetry
stored in the Standard Model lepton sector can be partially converted into a baryon asymmetry before electroweak sphaleron freeze-out. The
different regimes require kinetic treatments appropriate to the
relevant flavor and RHN dynamics.

Throughout this paper, the baryon asymmetry is evaluated numerically with
ULYSSES~\cite{Granelli:2020pim,Granelli:2026goh}. We use the two-RHN
density-matrix treatment (2DME) for the non-degenerate regime
~\cite{Blanchet:2011xq}, the resonant treatment (2RES) for
quasi-degenerate RHNs at the TeV scale, and the BEARS\_3RHN solver
for the GeV-scale regime. The Dirac phase \(\delta\) is scanned
throughout, and both vanishing and thermal initial RHN abundances are
considered. In the figures, these two initial conditions are denoted by
\(R_N^{\rm ini}=0\) and \(R_N^{\rm ini}=1\), respectively.

We characterize the result by
\begin{equation}
\mathcal R_B
\equiv
\frac{\eta_B}{\eta_B^{\rm obs}},
\qquad
\eta_B^{\rm obs}
\simeq
6.10\times10^{-10},
\label{eq:RB}
\end{equation}
where
\(\eta_B=(n_B-n_{\bar B})/n_\gamma\), and the observed value is
inferred from cosmological measurements~\cite{Planck:2018vyg}.
Successful leptogenesis corresponds to \(\mathcal R_B=1\).

\begin{figure}[!htbp]
\centering
\includegraphics[width=0.99\textwidth]
{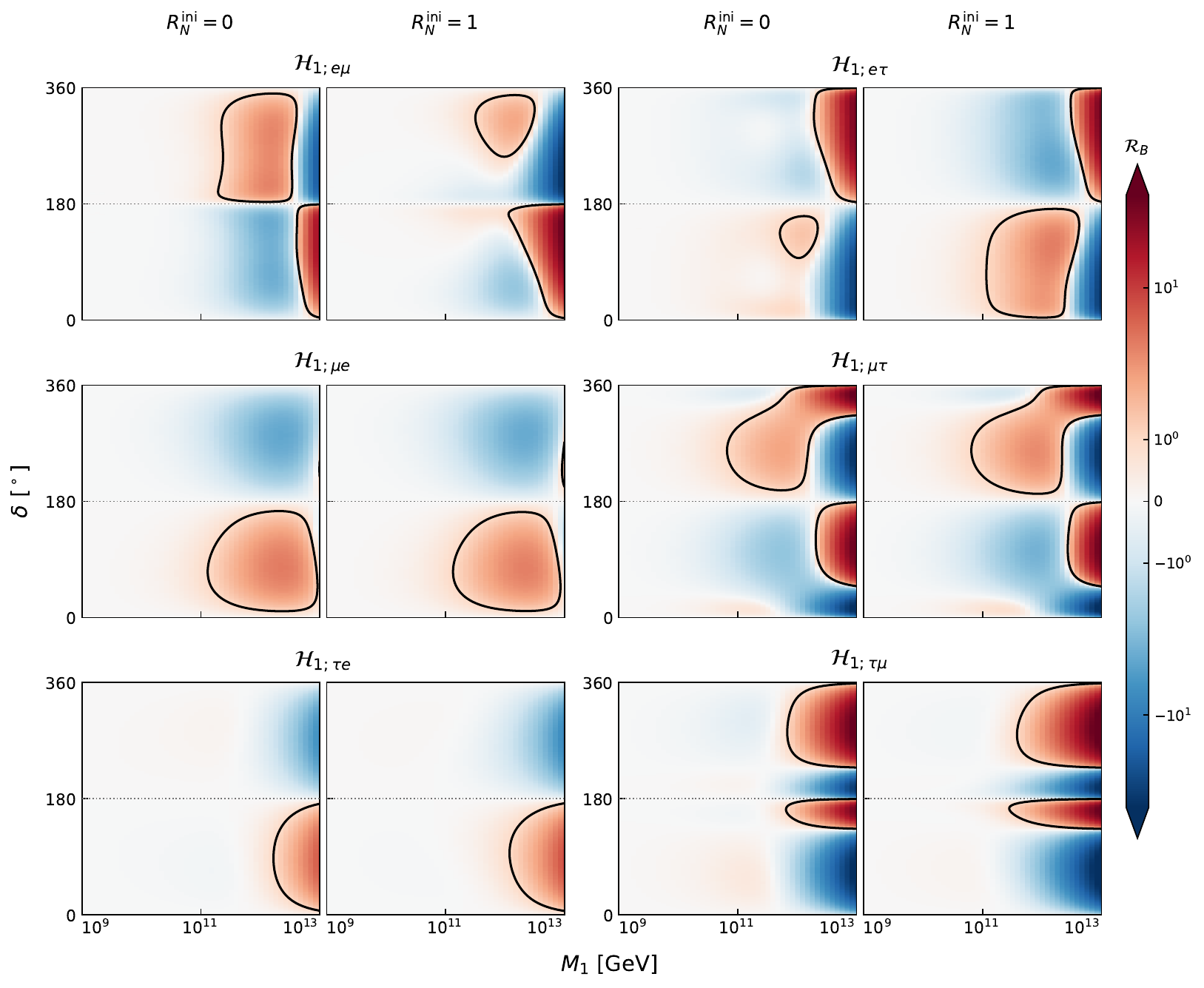}
\caption{
Non-degenerate leptogenesis in the explicit-CP realization
with \(M_2/M_1=3\) as a benchmark value, for the six
\(\mathcal H_{1;\alpha\beta}\) configurations.
Within each pair, the left and right panels correspond to vanishing
and thermal initial RHN abundances, respectively.
Colors show \(\mathcal R_B=\eta_B/\eta_B^{\rm obs}\) on a symmetric
logarithmic scale, and the solid black curves mark
\(\mathcal R_B=1\). The horizontal dotted lines indicate
\(\delta=180^\circ\). Figures~\ref{fig:main_nondegenerate_H1}
and~\ref{fig:main_nondegenerate_H2} share the same color scale.
}
\label{fig:main_nondegenerate_H1}
\end{figure}

\begin{figure}[!htbp]
\centering
\includegraphics[width=0.99\textwidth]
{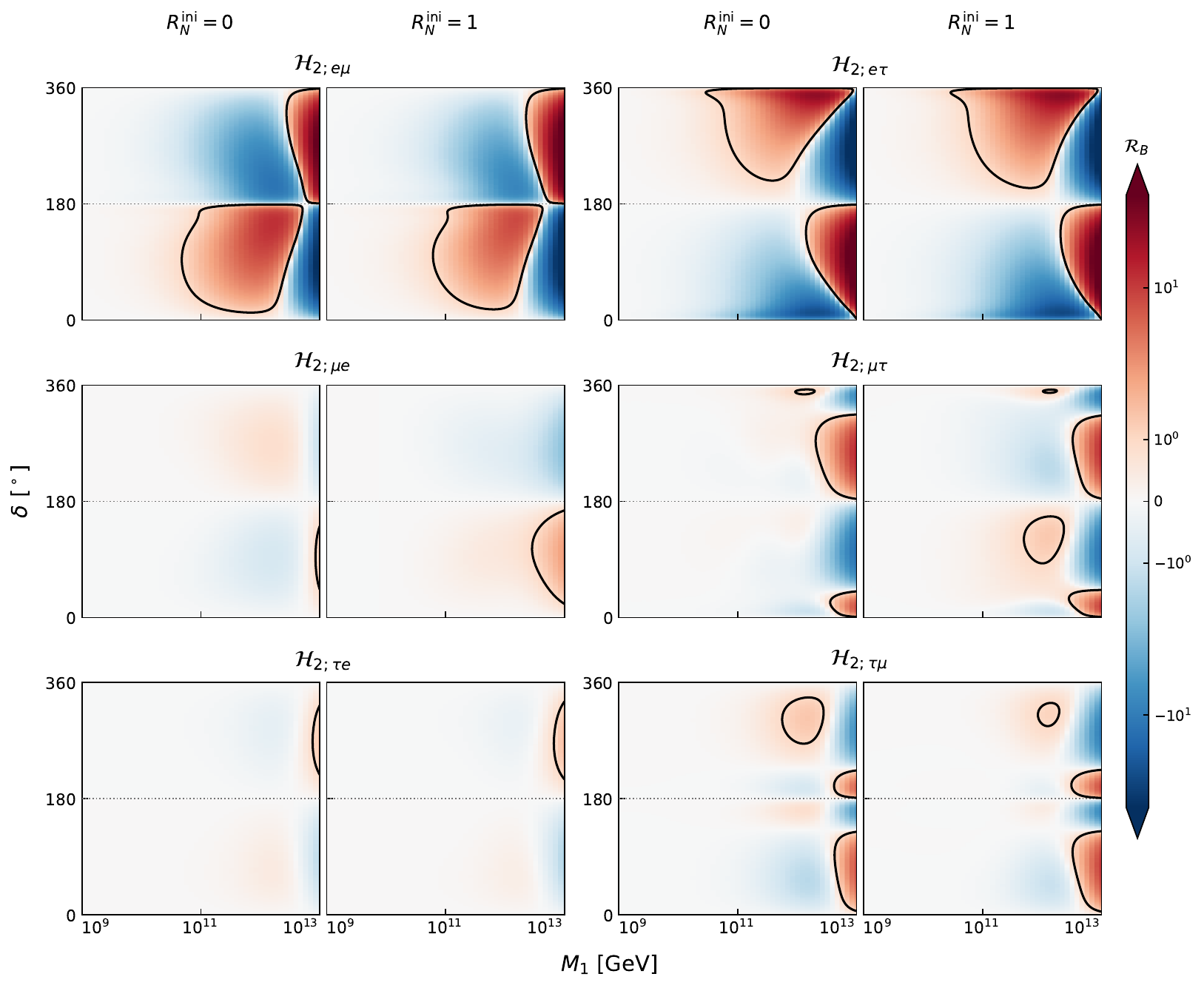}
\caption{
Same as Fig.~\ref{fig:main_nondegenerate_H1}, but for the six
\(\mathcal H_{2;\alpha\beta}\) configurations.
}
\label{fig:main_nondegenerate_H2}
\end{figure}

For the non-degenerate analysis, we take \(M_2/M_1=3\)
as a representative benchmark. For hierarchical RHN masses,
the results are largely insensitive to this ratio when
\(N_1\) dominates leptogenesis. We scan
\(10^9~\mathrm{GeV}\leq M_1\leq10^{13}~\mathrm{GeV}\)
together with the full range of the Dirac phase \(\delta\).
For both the explicit-CP and remnant-CP realizations, we restrict
the non-degenerate analysis to this mass range, since additional
washout from off-shell RHN-mediated \(\Delta L=2\) scatterings,
omitted in the present treatment, can become important
at higher scales. The results for the six
\(\mathcal H_{1;\alpha\beta}\) and six
\(\mathcal H_{2;\alpha\beta}\) configurations are displayed in
Figs.~\ref{fig:main_nondegenerate_H1} and
\ref{fig:main_nondegenerate_H2}.

For all twelve Yukawa configurations, regions satisfying
\(\mathcal R_B=1\) are found within the scanned parameter space for
both initial conditions. At the common CP-conserving solution
\(\delta=180^\circ\), the CP-violating source vanishes and no baryon
asymmetry is generated. Away from this point, the Yukawa couplings
vary with \(\delta\), producing a pronounced dependence of both the
sign and magnitude of \(\mathcal R_B\) on the Dirac phase and the
RHN mass scale.

The locations and shapes of the successful regions vary substantially
among the twelve configurations. In particular, configurations that
obey the same low-energy phase relation can lead to different
\(\mathcal R_B=1\) contours once the positions of the Yukawa zero
and the CP-sourcing Yukawa phase
are specified. The non-degenerate
leptogenesis results therefore distinguish high-energy Yukawa
configurations that are indistinguishable from their low-energy phase
correlations alone.

The dependence on the initial RHN abundance is also
configuration-dependent. For several configurations, the vanishing-
and thermal-abundance contours are nearly coincident, indicating that
the subsequent production and washout evolution largely erases the
initial-condition dependence. For other configurations, the contours
are displaced or contain different disconnected regions. The initial
RHN population therefore remains relevant in parts of the scanned
parameter space.

We next consider a quasi-degenerate RHN spectrum,
\(M_2\simeq M_1\). In this regime, the small RHN mass splitting and
the quantum coherence between the two states play an essential role
~\cite{Dev:2017wwc}. At the TeV scale, the enhancement is predominantly
associated with the mixing of quasi-degenerate RHNs in the resonant
regime. As the RHN mass scale is further lowered, coherent oscillations between the RHN states become increasingly important, and the dynamics gradually approaches the regime characteristic of ARS leptogenesis~\cite{Akhmedov:1998qx}.

For the quasi-degenerate leptogenesis analysis, we define
\begin{equation}
d_M
\equiv
\frac{M_2-M_1}{M_1},
\qquad
\Delta M=M_2-M_1,
\label{eq:dM}
\end{equation}
and introduce the average decay width
\begin{equation}
\overline{\Gamma}
\equiv
\frac{\Gamma_1+\Gamma_2}{2}.
\end{equation}
The self-energy contribution to the CP asymmetry can be resonantly
enhanced when the mass splitting becomes comparable to the decay-width
scale~\cite{Pilaftsis:2003gt}. We therefore study the viable parameter space on the
\((d_M,\delta)\) plane, taking \(M_1=1~\mathrm{TeV}\) as a
benchmark value for the TeV-scale resonant regime.

\begin{figure}[!htbp]
\centering
\includegraphics[width=0.85\textwidth]{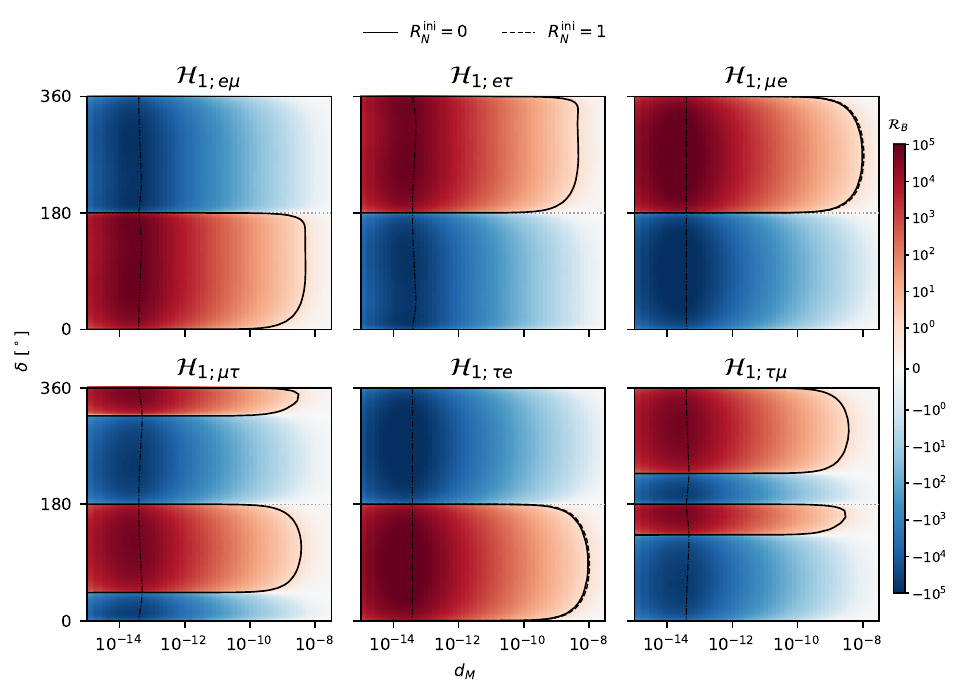}
\caption{
Resonant leptogenesis at \(M_1=1~\mathrm{TeV}\) for the six
\(\mathcal H_{1;\alpha\beta}\) configurations with the texture zero in
the first RHN column. The colored regions show
\(\mathcal R_B=\eta_B/\eta_B^{\rm obs}\) for vanishing initial RHN
abundance in the \((d_M,\delta)\) plane. The solid black contours mark
\(\mathcal R_B=1\) for this initial condition, while the dashed black
contours show the corresponding results for thermal initial RHN
abundance where they are distinct. The dot-dashed curves mark
\(\Delta M/\overline{\Gamma}=1\), and the horizontal dotted lines
indicate \(\delta=\pi\).
}
\label{fig:H1_resonant}
\end{figure}

The two sets of configurations,
\(\mathcal H_{1;\alpha\beta}\) and \(\mathcal H_{2;\alpha\beta}\), are not independent in the
quasi-degenerate regime. Moving the texture zero from one RHN column to
the other is closely related to interchanging the roles of the two
RHNs. Since their masses are nearly equal, this interchange leaves
the production and washout dynamics approximately unchanged while
reversing the sign of the generated asymmetry for a given value of
\(\delta\). CP conjugation also reverses the sign of the asymmetry
and maps the phase to its reflection about \(\delta=\pi\). Combining
these two observations, the \(\mathcal H_{2;\alpha\beta}\) results
are approximately obtained by reflecting the corresponding
\(\mathcal H_{1;\alpha\beta}\) results about \(\delta=\pi\).
We have verified this correspondence numerically at matched scan
points for all six pairs of configurations and for both initial
conditions. We therefore display only the results for the 
\(\mathcal H_{1;\alpha\beta}\) configurations in
Fig.~\ref{fig:H1_resonant}.

The asymmetry vanishes at the common
CP-conserving solution \(\delta=\pi\), whereas away from this point
both its magnitude and sign depend strongly on \(\delta\) and \(d_M\).
The locations and shapes of the successful
\(\mathcal R_B=1\) contours also vary among the six displayed Yukawa
configurations. In particular, configurations that obey the same
low-energy phase relation can select different RHN mass splittings
and Dirac-phase intervals. The TeV-scale leptogenesis results
therefore distinguish high-energy Yukawa configurations that are
equivalent at the level of low-energy neutrino observables.

The approximate reflection relations between the
$\mathcal H_{1;\alpha\beta}$ and $\mathcal H_{2;\alpha\beta}$
configurations discussed above are specific to the
quasi-degenerate TeV-scale treatment and do not accurately describe
the full low-mass kinetic evolution obtained with BEARS. We therefore
retain both RHN-column placements for the results of \(M_1=100~\mathrm{GeV}\) and
\(1~\mathrm{GeV}\).

\begin{figure}[!htbp]
\centering
\includegraphics[width=0.98\textwidth]{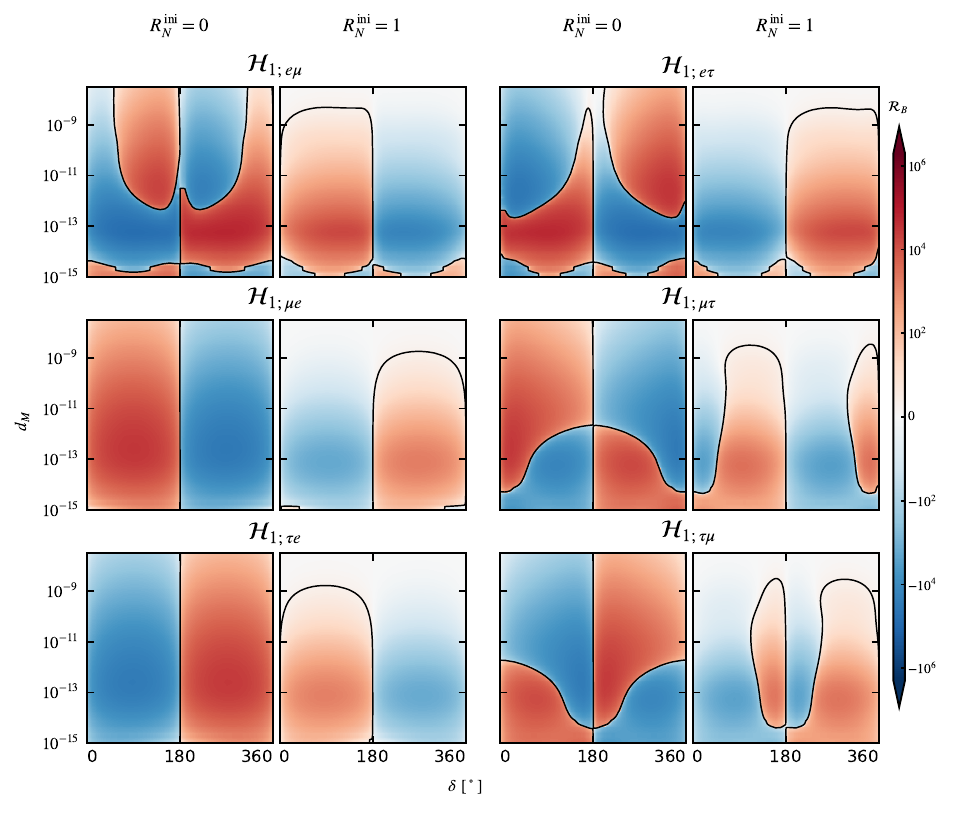}
\caption{
Low-scale leptogenesis at \(M_1=100~\mathrm{GeV}\) for the six
\(\mathcal H_{1;\alpha\beta}\) configurations of the explicit-CP
realization.
Within each pair, the left and right panels correspond to
vanishing and thermal initial RHN abundances, respectively.
Colors show \(\mathcal R_B=\eta_B/\eta_B^{\rm obs}\), and black
contours denote \(\mathcal R_B=1\) for the initial condition
displayed in that panel.
Both figures at this mass share the same color scale.
}
\label{fig:lowmass_100GeV_H1}
\end{figure}

\begin{figure}[!htbp]
\centering
\includegraphics[width=0.98\textwidth]{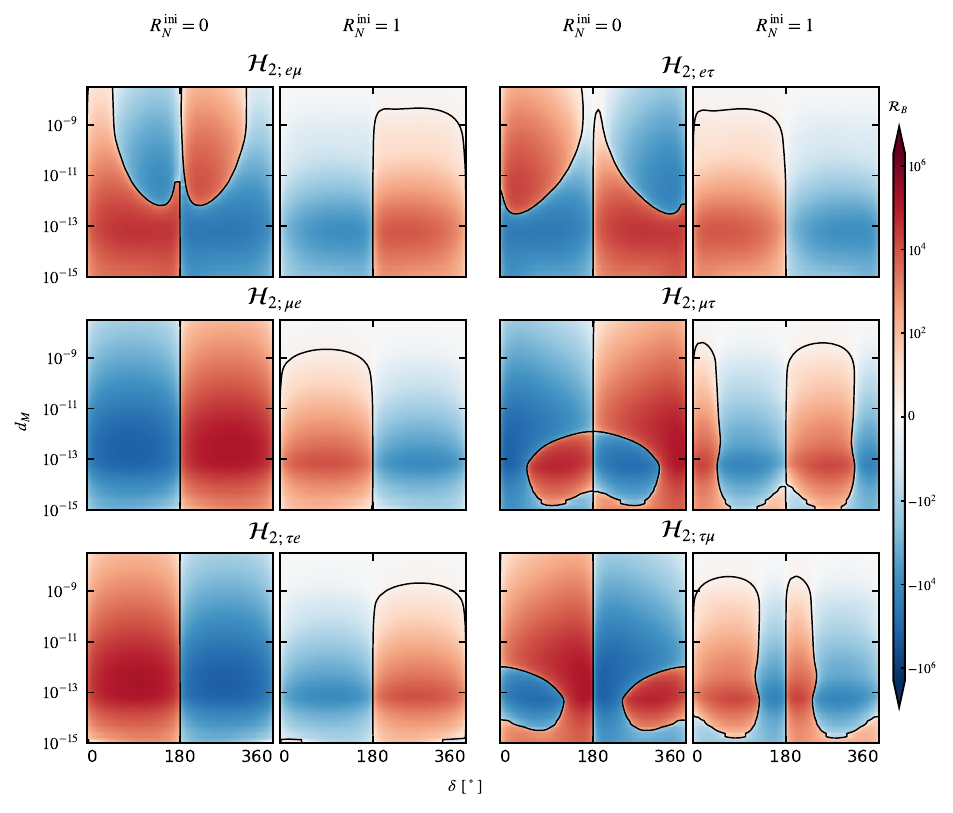}
\caption{
Same as Fig.~\ref{fig:lowmass_100GeV_H1}, but for the six
\(\mathcal H_{2;\alpha\beta}\) configurations.
}
\label{fig:lowmass_100GeV_H2}
\end{figure}

At \(M_1=100~\mathrm{GeV}\), successful leptogenesis remains possible.
Figures~\ref{fig:lowmass_100GeV_H1} and
\ref{fig:lowmass_100GeV_H2} show the six configurations with the
texture zero in the first and second RHN columns, respectively.

For all twelve configurations, regions satisfying
\(\mathcal R_B=1\) are present within the scanned parameter space
for both vanishing and thermal initial RHN abundances.
The locations and shapes of the successful contours can differ
substantially between the two initial conditions, indicating that
the asymmetry at this mass scale becomes sensitive to the initial
RHN population. In particular, configurations that obey the
same low-energy phase relation may require different RHN mass
splittings to reproduce the observed asymmetry.

\begin{figure}[!htbp]
\centering
\includegraphics[width=0.98\textwidth]
{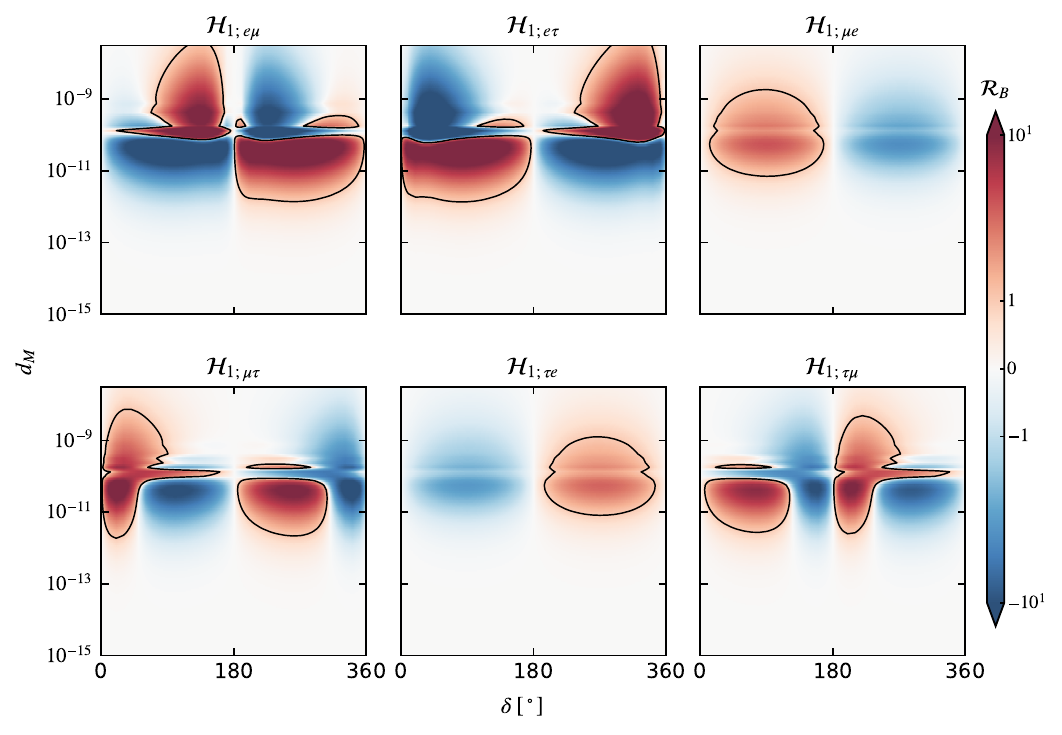}
\caption{
Low-scale leptogenesis at \(M_1=1~\mathrm{GeV}\) for the six
\(\mathcal H_{1;\alpha\beta}\) configurations.
The colored regions show
\(\mathcal R_B=\eta_B/\eta_B^{\rm obs}\)
in the \((d_M,\delta)\) plane for vanishing initial RHN abundance,
and the black contours denote \(\mathcal R_B=1\).
}
\label{fig:lowmass_1GeV_H1}
\end{figure}

\begin{figure}[!htbp]
\centering
\includegraphics[width=0.98\textwidth]
{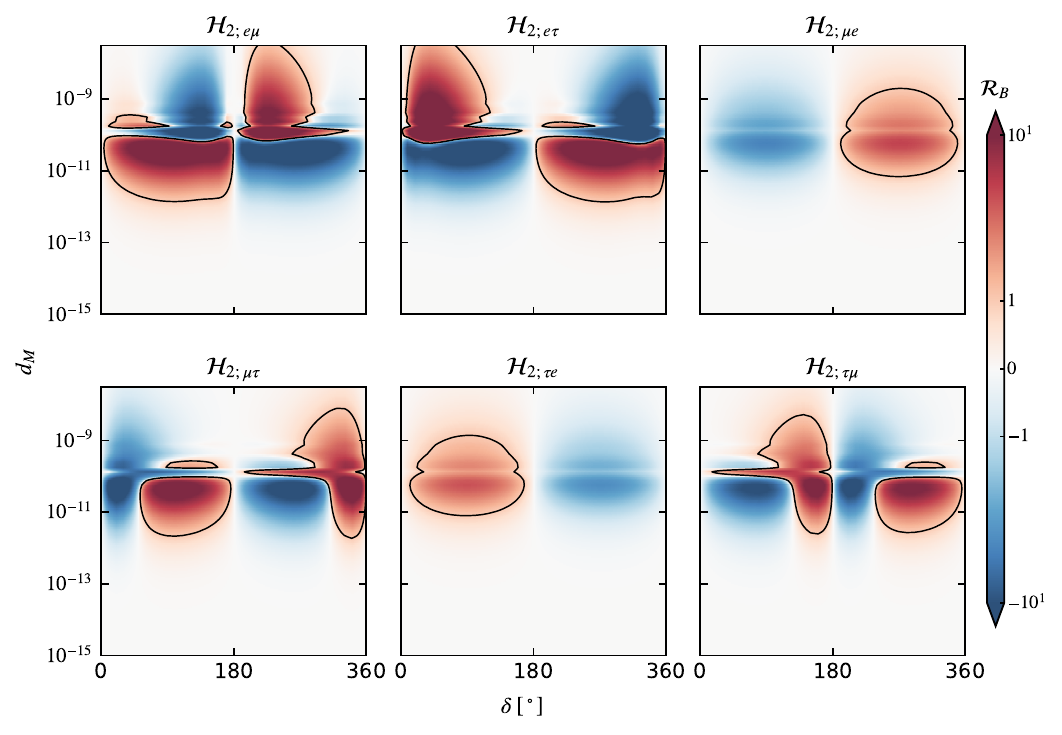}
\caption{
Same as Fig.~\ref{fig:lowmass_1GeV_H1}, but for the six
\(\mathcal H_{2;\alpha\beta}\) configurations.
}
\label{fig:lowmass_1GeV_H2}
\end{figure}

The situation changes qualitatively for the results of
\(M_1=1~\mathrm{GeV}\). The corresponding results are shown in
Figs.~\ref{fig:lowmass_1GeV_H1} and
\ref{fig:lowmass_1GeV_H2}.

The result exhibits a pronounced dependence
on the initial RHN abundance. For vanishing initial abundance, all
twelve Yukawa configurations contain regions in the scanned
\((d_M,\delta)\) plane where \(\mathcal R_B=1\) holds, although the
locations and structures of the successful contours vary significantly
among the different configurations. By contrast, for thermal initial
abundance, no resolved \(\mathcal R_B=1\) contour is found within the
same scanned parameter space. This is because the dominant ARS contribution is
generated during the out-of-equilibrium production and coherent
oscillations of the RHNs. For an initially thermal RHN population,
these processes are suppressed.

Taken together, the results show that the explicit-CP realization
admits successful leptogenesis in all the RHN mass regimes considered here. At \(M_1=100~\mathrm{GeV}\), viable
solutions are found for both vanishing and thermal initial RHN
abundances, whereas at \(M_1=1~\mathrm{GeV}\) successful regions are
obtained for vanishing initial abundance but no resolved successful
contour is found for thermal initial abundance. The low-energy phase relation alone therefore does
not determine the viability of leptogenesis: the absolute RHN mass
scale, the degree of mass degeneracy, the initial RHN abundance, and
the detailed high-energy Yukawa configuration all play essential
roles.

We finally assess how successful leptogenesis restricts the
approximate-zero behavior. Considering only the allowed values of the Dirac phase \(\delta\),
we find that the best attainable fixed-position suppression
remains close to its unrestricted value, with the largest relative
shift among the successful scans being about \(12\%\).
Successful leptogenesis therefore narrows the allowed phase regions
without spoiling the approximate-zero behavior.

\section{Remnant-CP MMFGY}
\label{sec:remnant_cp}

We now turn to the remnant-CP realization. The combination of one Yukawa texture zero with a remnant CP symmetry in the minimal seesaw has been studied in Ref.~\cite{Barreiros:2018bju}, which identified the viable configurations and analyzed their predictions for the low-energy CP phases and the effective Majorana mass governing neutrinoless double-beta decay.

Based on this framework, we investigate the approximate-two-zero behavior and leptogenesis of the viable normal-ordering configurations. We first examine how strongly a second Yukawa entry can be suppressed compared to other nonzero entries while maintaining the texture zero and remnant CP constraints. We then study whether these configurations can reproduce the observed baryon asymmetry in the different RHN mass regimes as considered in Sec.~2, and how successful leptogenesis restricts the approximate-two-zero behavior.

\subsection{Theoretical framework}
\label{sec:remnant_cp_theory}

Under generalized CP transformations, the left- and right-handed neutrino fields transform as
\begin{equation}
  \nu_L
  \rightarrow
  iX_\nu\gamma^0\nu_L^c,
  \qquad
  N_R
  \rightarrow
  iX_R\gamma^0N_R^c ,
\end{equation}
where \(X_\nu\) and \(X_R\) are unitary matrices acting in the
left- and right-handed neutrino flavor spaces, respectively.
Invariance of the neutrino Yukawa and Majorana mass terms imposes
the conditions~\cite{Barreiros:2018bju}
\begin{equation}
  X_\nu^\dagger Y X_R=Y^*,
  \qquad
  X_R^T M_R X_R=M_R^* .
  \label{eq:remnant_cp_conditions}
\end{equation}

It is convenient to express the CP transformations in the neutrino
mass bases through
\begin{equation}
  \widehat X_\nu
  \equiv
  U^\dagger X_\nu U^*,
  \qquad
  \widehat X_R
  \equiv
  U_R^\dagger X_R U_R^*,
\end{equation}
where \(U_R\) diagonalizes the RHN mass matrix according to
\(U_R^T M_R U_R=D_N\). In the diagonal-\(M_R\) basis adopted here,
\(U_R=\mathbf 1\). Following Ref.~\cite{Barreiros:2018bju}, we denote
the Casas--Ibarra matrix by \(O\) in this section. We retain the
convention of Eq.~\eqref{eq:CI}, namely,
\begin{equation}
  Y=\frac{i}{v}UD_\nu^{1/2}OD_N^{1/2},
  \qquad O\equiv R .
\end{equation}
With this convention, the remnant CP conditions imply
\begin{equation}
  O=-\widehat X_\nu O^*\widehat X_R^\dagger .
  \label{eq:remnant_cp_O}
\end{equation}
The overall minus sign originates from the explicit factor
\(i\) in the Casas--Ibarra parametrization.

The nonzero elements of \(O\) are
consequently restricted to be real or purely imaginary.
Together with orthogonality, this reduces the general complex
angle to three one-dimensional branches.

For normal ordering, representative forms of $O$ are~\cite{Barreiros:2018bju}
\begin{align}
  O_{\rm I}
  &=
  \begin{pmatrix}
    0 & 0\\
    \cos z & -\sin z\\
    \xi\sin z & \xi\cos z
  \end{pmatrix},
  \nonumber\\[1mm]
  O_{\rm II}
  &=
  \begin{pmatrix}
    0 & 0\\
    \cosh z & -i\sinh z\\
    i\xi\sinh z & \xi\cosh z
  \end{pmatrix},
  \nonumber\\[1mm]
  O_{\rm III}
  &=
  \begin{pmatrix}
    0 & 0\\
    i\sinh z & -\cosh z\\
    \xi\cosh z & i\xi\sinh z
  \end{pmatrix},
  \qquad
  z\in\mathbb{R},
  \label{eq:remnant_cp_orbits}
\end{align}
with $\xi=\pm1$. Equivalently, these branches may be represented by
\begin{equation}
  \widehat z=z,
  \qquad
  \widehat z=iz,
  \qquad
  \widehat z=\frac{\pi}{2}-iz,
\end{equation}
up to discrete shifts and sign conventions.

The \(O_{\rm I}\) branch connects this framework to leptogenesis
driven by low-energy leptonic CP phases, as studied by assuming
a real Casas--Ibarra matrix
~\cite{Pascoli:2006ie,Pascoli:2006ci}.
On this branch, \(O\) introduces no independent
CP-violating phase, and the CP violation relevant for
leptogenesis entirely originates from the phases in \(U\).
Here the real-\(O\) structure arises as an allowed branch
of the remnant CP constraint.
Based on this, we then impose a single Yukawa texture zero. The six possible
positions are
\begin{equation}
\begin{aligned}
  Y_1:\quad &Y_{e1}=0,   &\qquad
  Y_2:\quad &Y_{\mu1}=0, &\qquad
  Y_3:\quad &Y_{\tau1}=0,\\
  Y_4:\quad &Y_{e2}=0,   &\qquad
  Y_5:\quad &Y_{\mu2}=0, &\qquad
  Y_6:\quad &Y_{\tau2}=0.
\end{aligned}
\label{eq:remnant_cp_textures}
\end{equation}
The texture-zero condition fixes the remaining real parameter $z$ on
each allowed CP branch and simultaneously imposes a relation among the
low-energy CP phases.

For normal ordering, the allowed texture--branch combinations in the
setup considered here are
\begin{equation}
\begin{aligned}
  Y_1 &: \quad O_{\rm I},\,O_{\rm III}; \quad &
  Y_4 &: \quad O_{\rm I},\,O_{\rm II};\\
  Y_2 &: \quad O_{\rm I},\,O_{\rm II}; &
  Y_5 &: \quad O_{\rm I},\,O_{\rm III};\\
  Y_3 &: \quad O_{\rm I},\,O_{\rm II};&
  Y_6 &: \quad O_{\rm I},\,O_{\rm III}.
\end{aligned}
\label{eq:remnant_cp_allowed_cases}
\end{equation}

Under interchange of the two RHN columns, one arrives at
\begin{equation}
  Y_1\leftrightarrow Y_4,\qquad
  Y_2\leftrightarrow Y_5,\qquad
  Y_3\leftrightarrow Y_6 .
\end{equation}
At the same time, the \(O_{\rm I}\) branch is mapped within
itself, whereas the \(O_{\rm II}\) and \(O_{\rm III}\) branches are
interchanged, up to the discrete sign and parameter conventions in
Eq.~\eqref{eq:remnant_cp_orbits}.

These twelve combinations define the parameter space considered below. Their low-energy phase relations and neutrinoless-double-beta-decay
predictions have already been studied in
Ref.~\cite{Barreiros:2018bju}.

\subsection{Phenomenological results}
\label{sec:remnant_cp_pheno}

The established low-energy phase correlations and
neutrinoless-double-beta-decay predictions of the remnant-CP
realization given in Ref.~\cite{Barreiros:2018bju} are not repeated here. As a complementary study, we focus on the same
two high-energy questions studied for the explicit-CP realization:
whether an approximate second Yukawa zero can emerge and how the
resulting Yukawa structures affect leptogenesis.

\subsubsection{Approximate zeros}
\label{sec:remnant_cp_approx_zero}

We first examine whether the remnant-CP and one-zero conditions can
allow for an approximate second zero in the Yukawa matrix. To facilitate
a direct comparison with the explicit-CP realization, we use the
same suppression measure \(\epsilon_{\rho j}\) defined in
Eq.~\eqref{eq:second_zero_ratio}. \(\epsilon_{\rho j}\ll1\) indicates
that the candidate entry is strongly suppressed relative to every other nonzero
entry in the same RHN column. As before, the common factor
\(\sqrt{M_j}\) cancels, making the suppression measure independent of
the overall RHN mass scale.

For each allowed texture--branch combination, we determine the
candidate position \((\rho_*,j_*)\) whose suppression curve reaches the
smallest value over the full \(\delta\) scan. This position is selected
once and then held fixed for all values of \(\delta\). We therefore
plot
\begin{equation}
  \epsilon_{\rm cand}(\delta)
  \equiv
  \epsilon_{\rho_*j_*}(\delta),
  \label{eq:remnant_cp_candidate_epsilon}
\end{equation}
rather than reselecting the smallest Yukawa entry at every phase point.
The latter procedure would combine different candidate positions into
an artificial lower envelope and would not correspond to a single
approximate texture. With the column-exchange correspondence described above, the paired
configurations give identical dimensionless suppression curves and are
therefore displayed together as \(Y_1/Y_4\), \(Y_2/Y_5\), and
\(Y_3/Y_6\).

\begin{figure}[!htbp]
  \centering
  \includegraphics[width=0.99\textwidth]{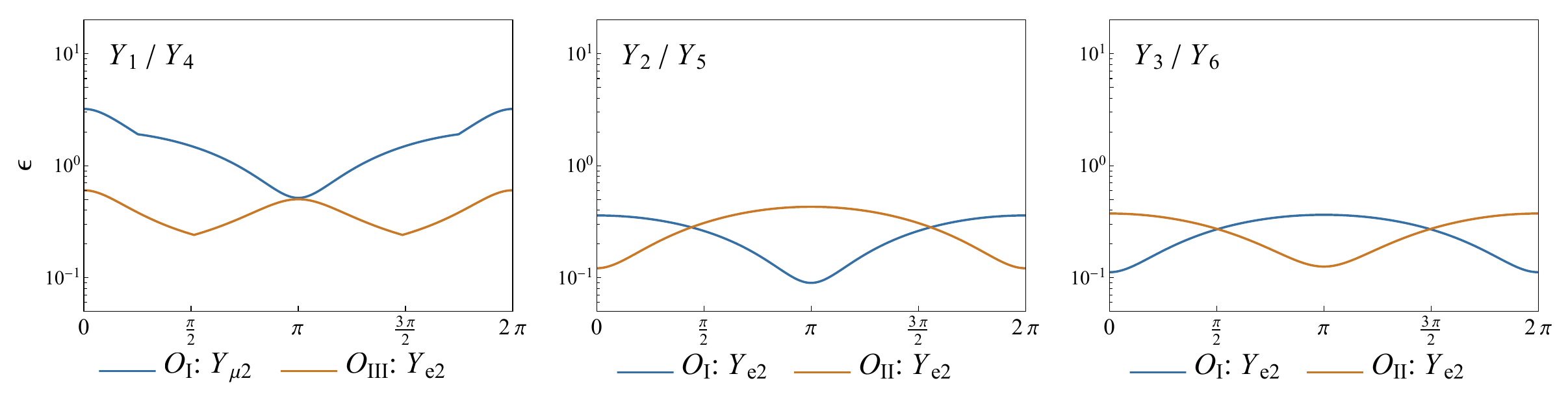}
  \caption{
    Fixed-position approximate-second-zero measure for the one-zero
    Yukawa textures subject to the remnant CP symmetry. For
    each texture--branch combination, the candidate entry shown in the
    legend is selected once from the full phase scan and then held fixed
    throughout the variation of \(\delta\). The vertical axis shows
    \(\epsilon_{\rm cand}\), defined in
    Eq.~\eqref{eq:remnant_cp_candidate_epsilon}, on a logarithmic
    scale. The column-exchange partners \(Y_1/Y_4\), \(Y_2/Y_5\), and
\(Y_3/Y_6\) have identical suppression curves, with \(O_{\rm I}\)
mapped within itself and \(O_{\rm II}\) and \(O_{\rm III}\)
interchanged under the column exchange.
  }
  \label{fig:remnant_cp_approx_zero}
\end{figure}

The strongest suppression occurs for the \(Y_2/Y_5\) class on the
\(O_{\rm I}\) branch, for which
\(\epsilon_{\rm cand}^{\rm min}\simeq0.090\) at
\(\delta=\pi\). The $Y_2$--$O_{\rm II}$ and $Y_5$--$O_{\rm III}$ configurations,
together with all four allowed configurations associated with
$Y_3$ and $Y_6$, reach
\(\epsilon_{\rm cand}^{\min}\simeq0.11\text{--}0.13\),
corresponding to an approximate one-order-of-magnitude hierarchy.
By contrast, the \(Y_1\)--\(O_{\rm III}\) and
\(Y_4\)--\(O_{\rm II}\) configurations reach only
\(\epsilon_{\rm cand}^{\rm min}\simeq0.24\), while their
\(O_{\rm I}\) counterparts reach
\(\epsilon_{\rm cand}^{\rm min}\simeq0.52\). These results show that selected texture--branch combinations
allow an approximate second zero at the level of a ten-percent
relative Yukawa coupling.

Applying a similar full-matrix comparison to the
remnant-CP realization gives
\(r_{\rm global}^{\min}\simeq0.20\), attained at \(\delta=\pi\)
for the \(Y_2\) and \(Y_5\) configurations on the \(O_{\rm I}\)
branch. This corresponds to a strongest full-matrix hierarchy
of approximately \(1:5\).

\subsubsection{Leptogenesis}

We next examine whether the remnant-CP framework can reproduce the
observed baryon asymmetry. To allow a direct comparison with the
explicit-CP realization, we use the same oscillation inputs,
RHN-mass benchmarks, numerical setup, and baryon-asymmetry measure
\(\mathcal{R}_B\) defined in Eq.~\eqref{eq:RB}.

For each value of the Dirac phase \(\delta\), the texture-zero and
remnant-CP conditions determine the remaining low-energy phase and
the Casas--Ibarra parameter.

For the allowed Casas--Ibarra branches, the flavor-summed decay
CP asymmetry vanishes, while individual flavor asymmetries can
remain nonzero~\cite{Li:2017zmk}. Charged-lepton flavor effects
are therefore essential for generating a net baryon asymmetry.
The scale \(10^{12}~\mathrm{GeV}\) does not constitute a sharp
upper boundary for these effects: density-matrix studies have
shown that finite charged-lepton-induced decoherence can sustain
a nonzero asymmetry above this scale even when the flavor-summed
CP asymmetry vanishes~\cite{Moffat:2018smo,
Granelli:2021fyc}. Our numerical analysis uses the
two-RHN density-matrix equations, which retain the continuous
evolution of flavor coherence rather than imposing an abrupt
transition at \(10^{12}~\mathrm{GeV}\).

\begin{figure}[!htbp]
  \centering
  \includegraphics[width=\textwidth]{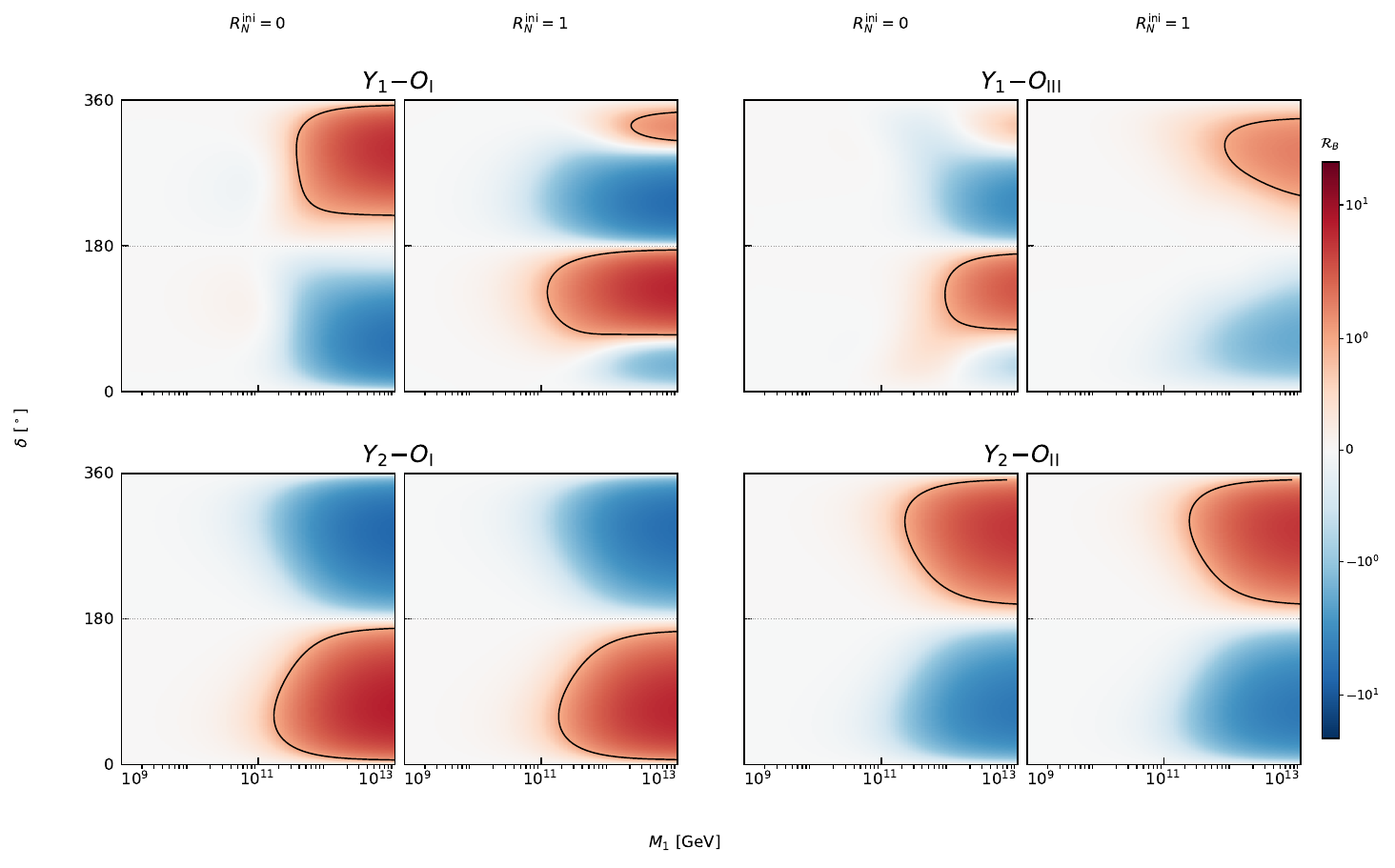}
  \caption{
    Non-degenerate leptogenesis in the remnant-CP framework
    with normal ordering and $M_2/M_1=3$, for the four configurations
    associated with $Y_1$ and $Y_2$.
    Within each pair, the left and right panels correspond to
    vanishing and thermal initial RHN abundances, respectively.
    Colors show $\mathcal R_B=\eta_B/\eta_B^{\rm obs}$, and
    solid black curves mark $\mathcal R_B=1$ for the initial
    condition displayed in that panel.
    Horizontal dotted lines indicate $\delta=180^\circ$.
    Both non-degenerate figures share the same color scale.
  }
  \label{fig:remnant_cp_nondegenerate}
\end{figure}

\begin{figure}[!htbp]
  \centering
  \includegraphics[width=\textwidth]{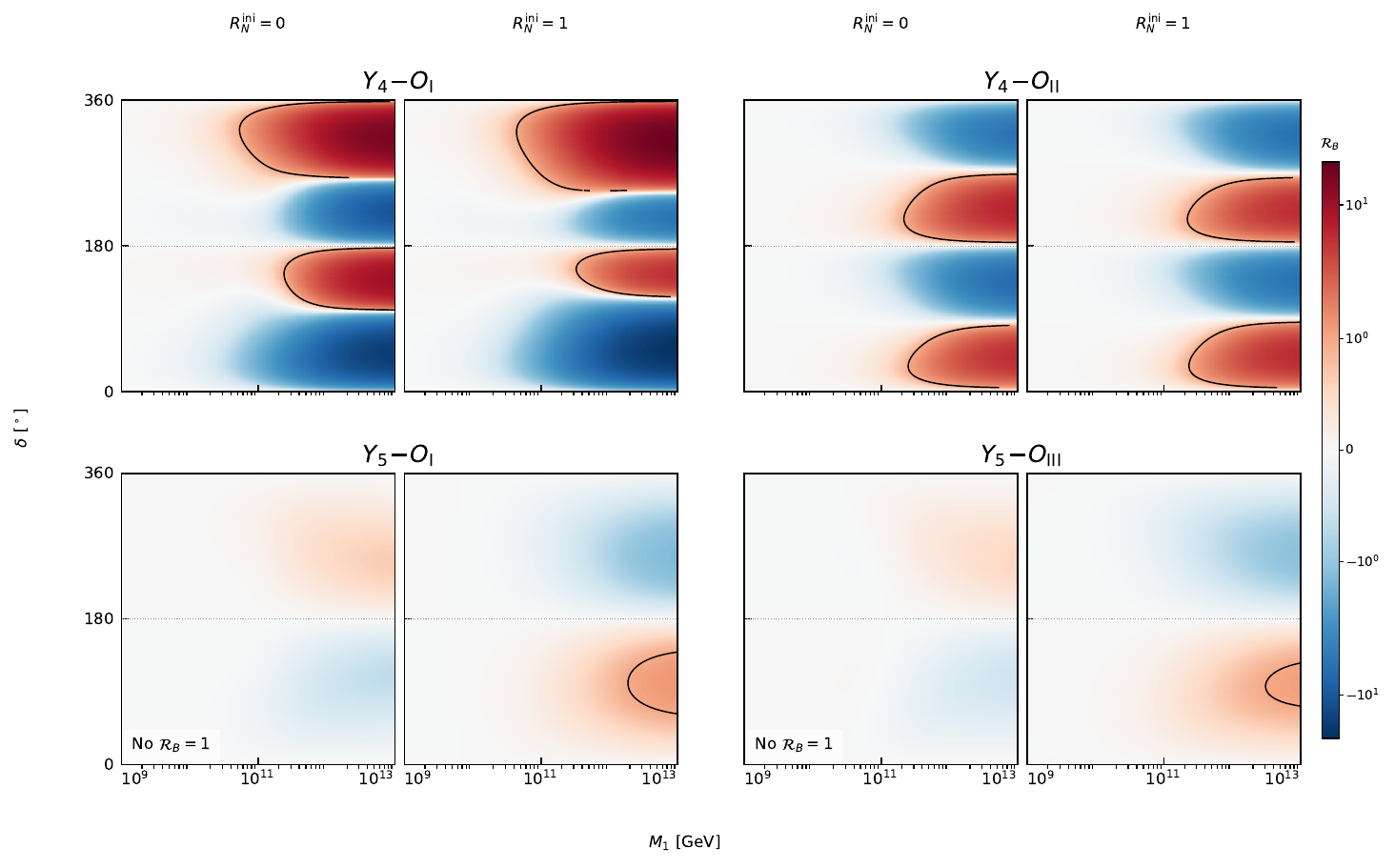}
  \caption{
    Same as Fig.~\ref{fig:remnant_cp_nondegenerate},
    but for the four configurations associated with $Y_4$ and $Y_5$.
  }
  \label{fig:remnant_cp_nondegenerate_Y4_Y5}
\end{figure}

For the non-degenerate analysis, we still take $M_2/M_1=3$ as a benchmark value and scan
$10^9~{\rm GeV}\le M_1\le10^{13}~{\rm GeV}$ together with the full
range of $\delta$. All twelve allowed texture--branch combinations are
first subjected to a common coarse scan. The eight combinations for which successful solutions are found
are then adaptively refined and displayed in
Figs.~\ref{fig:remnant_cp_nondegenerate}
and~\ref{fig:remnant_cp_nondegenerate_Y4_Y5}. The remaining four cases,
$Y_3$--$O_{\rm I}$, $Y_3$--$O_{\rm II}$,
$Y_6$--$O_{\rm I}$, and $Y_6$--$O_{\rm III}$, only yield
$\max\mathcal{R}_B<4\times10^{-2}$
for both initial conditions and are therefore not shown.

Several features are observed. First, both the sign and magnitude of
the baryon asymmetry depend strongly on \(\delta\). The asymmetry changes
sign across the special phase \(\delta=180^\circ\), around which the
remnant-CP phase relations approach CP-parity limits. Second, the
minimum RHN mass compatible with \(\mathcal{R}_B=1\) is highly
configuration dependent, lying broadly in the range
\(M_1\sim10^{10}\)--\(10^{12}~{\rm GeV}\). Comparing the paired panels
shows that the successful contours depend substantially on the initial
RHN abundance for some configurations, whereas others exhibit much
weaker initial-condition dependence. Finally, configurations related
by exchanging the RHN columns need not give equivalent leptogenesis
predictions once the mass ordering \(M_1<M_2\) is
fixed. The two columns are then associated with RHNs of different
masses and therefore enter the production and washout dynamics
differently.

\begin{figure}[!htbp]
  \centering
  \includegraphics[width=0.95\textwidth]{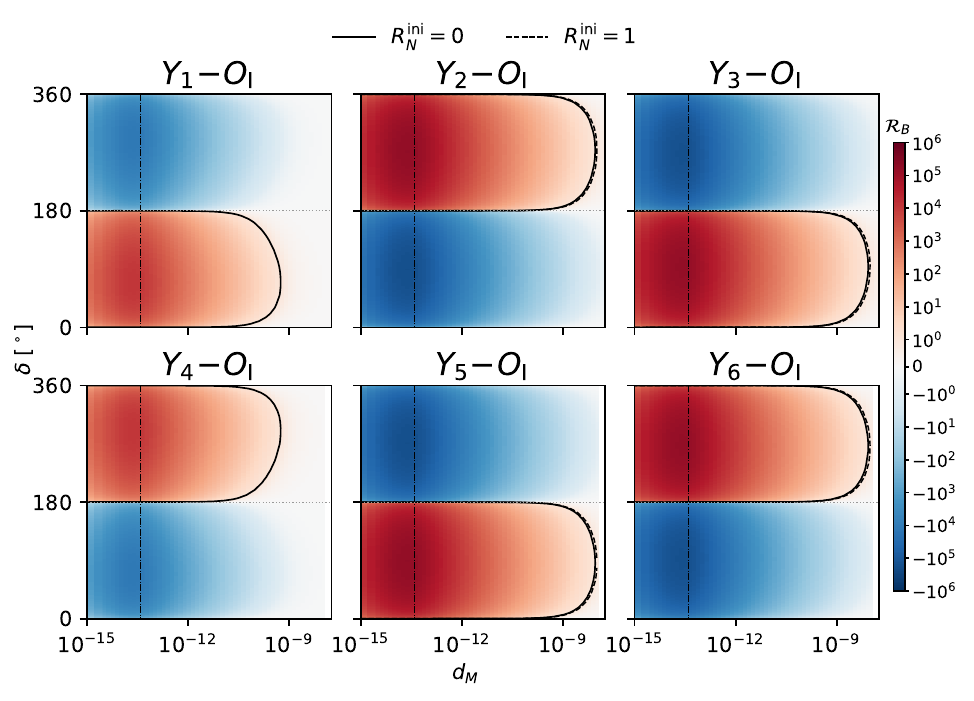}
  \caption{
    Resonant leptogenesis results at \(M_1=1~\mathrm{TeV}\) for the
    six remnant-CP configurations
    \(Y_i\)--\(O_{\rm I}\), with \(i=1,\ldots,6\).
    Colors show \(\mathcal R_B=\eta_B/\eta_B^{\rm obs}\)
    for vanishing initial RHN abundance.
    Solid black contours mark \(\mathcal R_B=1\) for
    vanishing initial abundance; dashed black contours
    show the thermal-initial-abundance results where
    they differ from the solid contours.
    Dot-dashed curves mark \(\Delta M/\overline{\Gamma}=1\),
    and horizontal dotted lines indicate \(\delta=180^\circ\).
  }
  \label{fig:remnant_cp_resonant}
\end{figure}

For the quasi-degenerate benchmark \(M_1=1~\mathrm{TeV}\),
we scan the relative mass splitting \(d_M\), defined in
Eq.~\eqref{eq:dM}, together with the Dirac phase \(\delta\).
All twelve remnant-CP Yukawa configurations are
evaluated for both vanishing and thermal initial RHN abundances.
Within the scanned parameter region and numerical resolution,
successful leptogenesis is found only for the
\(O_{\rm I}\) configurations.
For the six configurations associated with \(O_{\rm II}\)
or \(O_{\rm III}\), the resulting asymmetry satisfies
\(\lvert\mathcal R_B\rvert<4.2\times10^{-9}\) under both initial conditions.
These configurations are therefore omitted from the figure,
while the results for six \(O_{\rm I}\) configurations are collected in
Fig.~\ref{fig:remnant_cp_resonant}.

The six \(O_{\rm I}\) configurations select different ranges
of the Dirac phase for successful leptogenesis.
For \(Y_1\), \(Y_3\), and \(Y_5\), the successful contours
lie below \(\delta=\pi\), whereas those for
\(Y_2\), \(Y_4\), and \(Y_6\) lie above it.
In particular, each column-exchange pair,
\(Y_1/Y_4\), \(Y_2/Y_5\), and \(Y_3/Y_6\),
selects opposite sides of \(\delta=\pi\).
Thus, even though the paired configurations have identical
predictions for low-energy neutrino observables,
they favor different Dirac-phase intervals when successful
leptogenesis is required.

For the lower mass benchmarks \(M_1=100~\mathrm{GeV}\) and
\(1~\mathrm{GeV}\), we use the same numerical setup as for
the explicit-CP realization. All twelve remnant-CP Yukawa
configurations are analyzed for both vanishing and thermal initial RHN
abundances.

\begin{figure}[!htbp]
  \centering
  \includegraphics[width=\textwidth]{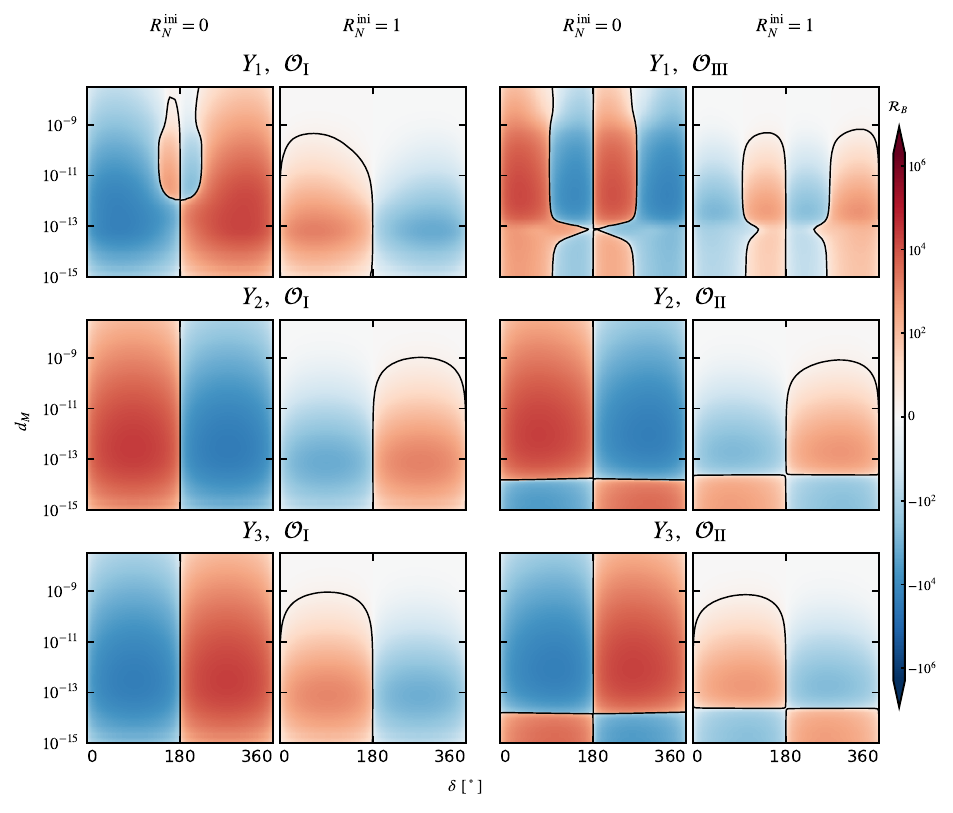}
  \caption{
    Low-scale leptogenesis at \(M_1=100~\mathrm{GeV}\)
    for the six remnant-CP configurations associated
    with \(Y_1\), \(Y_2\), and \(Y_3\).
    Within each pair, the left and right panels correspond to
    vanishing and thermal initial RHN abundances, respectively.
    Colors show \(\mathcal R_B=\eta_B/\eta_B^{\rm obs}\),
    and black contours mark \(\mathcal R_B=1\) for the
    initial condition displayed in that panel.
    Both figures at this mass scale share the same color scale.
  }
  \label{fig:rcp_100GeV_Y1_Y3}
\end{figure}

\begin{figure}[!htbp]
  \centering
  \includegraphics[width=\textwidth]{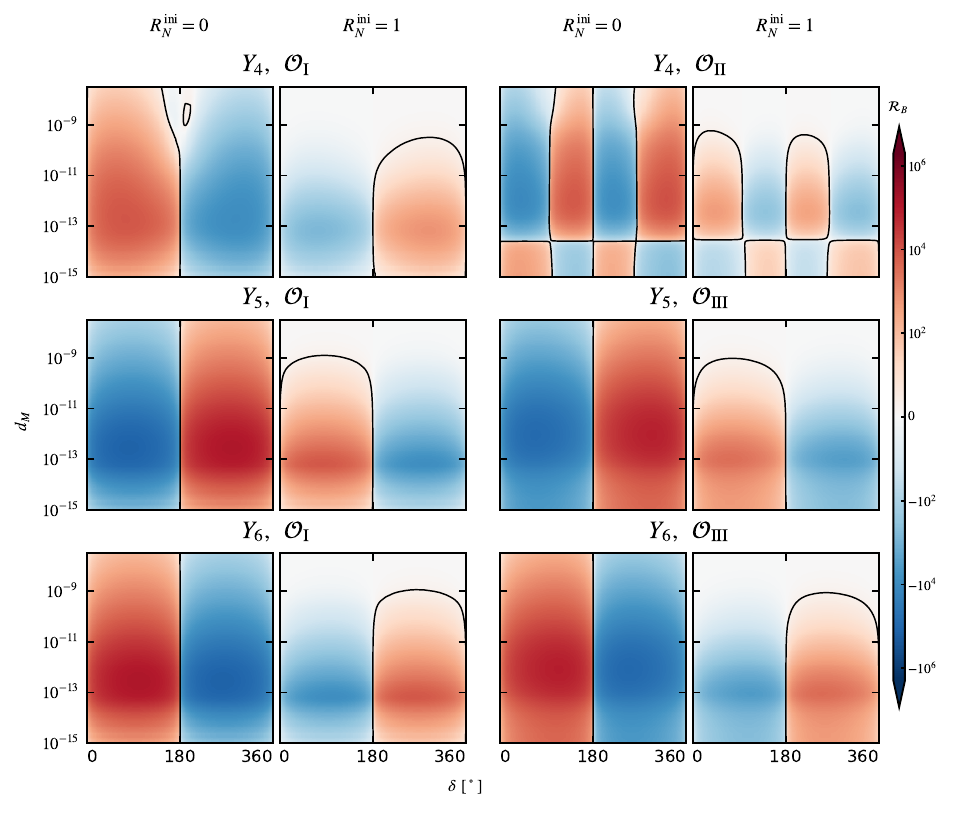}
  \caption{
    Same as Fig.~\ref{fig:rcp_100GeV_Y1_Y3},
    but for the six configurations associated with
    \(Y_4\), \(Y_5\), and \(Y_6\).
  }
  \label{fig:rcp_100GeV_Y4_Y6}
\end{figure}

The \(M_1=100~\mathrm{GeV}\) results are shown in
Figs.~\ref{fig:rcp_100GeV_Y1_Y3}
and~\ref{fig:rcp_100GeV_Y4_Y6}.
The two initial conditions are displayed in adjacent panels,
with each background and its black contour evaluated for the
same initial condition.

The results show that all twelve configurations admit
successful contours for both initial conditions, but the paired
panels reveal substantial differences in their predictions.
For example, the \(Y_1\)--\(O_{\rm I}\) configuration exhibits
a narrow successful branch at larger \(d_M\) for vanishing
initial abundance, whereas the thermal case favors smaller
splittings and a different range of \(\delta\).
The change is accompanied by a rearrangement of the positive
and negative asymmetry regions. Thus, the initial RHN population
affects more than the overall magnitude of the asymmetry:
it can affect what phase and mass-splitting combinations
are compatible with the observation.

The panels also distinguish the different Casas--Ibarra branches.
For a given \(Y_i\), the \(O_{\rm I}\) and
\(O_{\rm II/III}\) configurations can produce markedly different
contour shapes and sign distributions.
The low-scale leptogenesis therefore remains sensitive to the
high-energy Yukawa configuration.

\begin{figure}[!htbp]
  \centering
  \includegraphics[width=\textwidth]{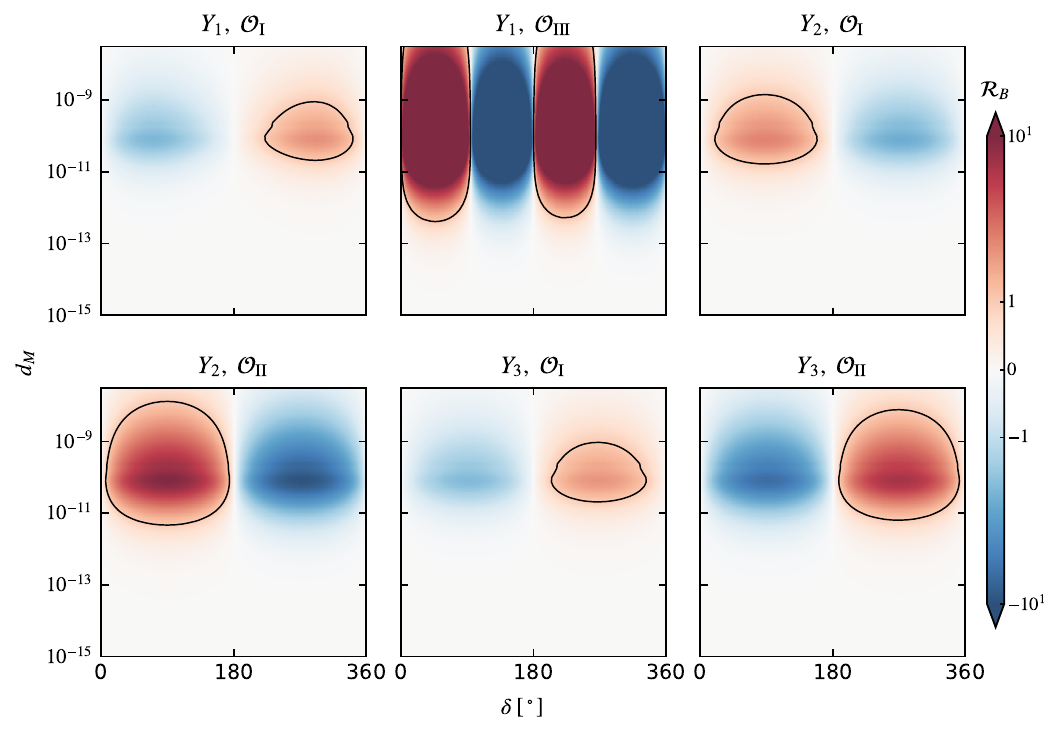}
  \caption{
    Low-scale leptogenesis at \(M_1=1~\mathrm{GeV}\)
    for the six remnant-CP configurations associated
    with \(Y_1\), \(Y_2\), and \(Y_3\).
    The colored background shows
    \(\mathcal R_B=\eta_B/\eta_B^{\rm obs}\)
    for vanishing initial RHN abundance, and solid black
    contours mark \(\mathcal R_B=1\).
    Both figures at this mass scale share the same color scale.
  }
  \label{fig:rcp_1GeV_Y1_Y3}
\end{figure}

\begin{figure}[!htbp]
  \centering
  \includegraphics[width=0.99\textwidth]{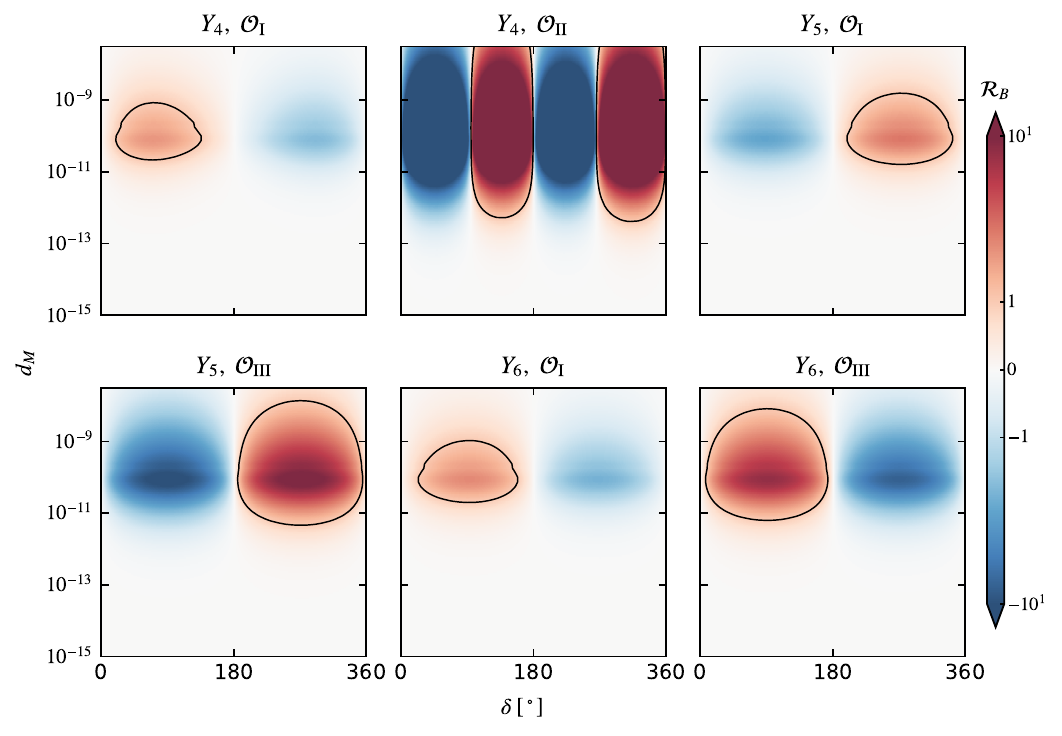}
  \caption{
    Same as Fig.~\ref{fig:rcp_1GeV_Y1_Y3},
    but for the six configurations associated with
    \(Y_4\), \(Y_5\), and \(Y_6\).
  }
  \label{fig:rcp_1GeV_Y4_Y6}
\end{figure}

For \(M_1=1~\mathrm{GeV}\), the dependence on the initial
RHN abundance is more restrictive.
Figures~\ref{fig:rcp_1GeV_Y1_Y3}
and~\ref{fig:rcp_1GeV_Y4_Y6} show the results for
vanishing initial abundance.

The successful regions for
vanishing initial abundance display a pronounced dependence
on the Casas--Ibarra branch.
The \(O_{\rm I}\) panels contain localized successful regions,
whereas the \(O_{\rm II/III}\) panels generally exhibit broader
structures with different extents in \(d_M\) and \(\delta\).

The role of the \(O_{\rm II}\) and \(O_{\rm III}\) branches
also changes between \(M_1=1~\mathrm{TeV}\) and
\(M_1=1~\mathrm{GeV}\).
Their asymmetries are strongly suppressed in the
\(1~\mathrm{TeV}\) resonant scan, while successful contours
appear at \(1~\mathrm{GeV}\) for vanishing initial abundance.
At the same time, no successful contour is resolved for
thermal initial abundance in the \(1~\mathrm{GeV}\) scan.
The viability of a configuration must therefore be assessed
together with the RHN mass regime and the assumed initial
population.

With the Dirac phase \(\delta\) restricted to values compatible with successful leptogenesis,
the best attainable fixed-position suppression is only moderately
weakened, with the largest relative shift among the scanned benchmarks
being about \(19\%\). Hence, successful leptogenesis restricts the allowed values
of the Dirac phase \(\delta\) but does not spoil
the approximate-zero behavior.


\section{Conclusions}
\label{sec:conclusion}

Current experimental results indicate a preference for normal
neutrino mass ordering. However, the highly predictive FGY ansatz,
which features two texture zeros in the neutrino Yukawa matrix
of the minimal seesaw model with two RHNs, is incompatible
with NO under current oscillation constraints.
In this work, we have therefore investigated two realizations
of the MMFGY framework. Both retain one Yukawa texture
zero but restrict the CP structure in different ways. In the explicit-CP realization, the two physical Yukawa phases
of a one-zero texture are examined separately as the sole source
of CP violation. In the
remnant-CP realization, the Casas--Ibarra freedom is instead constrained
by a remnant CP symmetry.

For the explicit-CP realization, the twelve distinct high-energy
Yukawa configurations reduce to three independent classes at low
energies, represented by the ordered flavor pairs
\((e,\tau)\), \((e,\mu)\), and \((\mu,e)\). Each class
predicts a correlation between the Dirac and Majorana phases. The corresponding predictions for the effective Majorana mass
relevant for neutrinoless double-beta decay are also investigated.
For the remnant-CP realization, the corresponding low-energy results have
already been presented in Ref.~\cite{Barreiros:2018bju}
and are not repeated here.

For both realizations, we have also examined the possible emergence
of an approximate second Yukawa zero. For some configurations,
a selected candidate entry can be suppressed to about one tenth
of every other nonzero entry in the same RHN column. In the
quasi-degenerate regime, the minimum ratio between the smallest
and second-smallest nonzero entries of the full Yukawa matrix
is about \(0.2\), corresponding to a hierarchy of approximately
\(1:5\). These structures can therefore be interpreted as
approximate remnants of the FGY hierarchy. Requiring successful
leptogenesis restricts the allowed parameter values and can weaken
the suppression, but does not qualitatively change this conclusion.

The leptogenesis analysis shows that the explicit-CP realization can
reproduce the observed baryon asymmetry in the non-degenerate
regime and in the quasi-degenerate regimes over the range of RHN
masses considered. In the TeV-scale quasi-degenerate regime, the \(\mathcal H_{1;\alpha\beta}\) and \(\mathcal H_{2;\alpha\beta}\) configurations are approximately related by an
interchange of the RHN columns together with a CP reflection. The
successful regions nevertheless remain sensitive to the flavor
assignment, the RHN mass spectrum, and, at the GeV scale, the
initial RHN abundance. The low-energy phase correlation alone is
therefore insufficient to determine the leptogenesis outcome.

The remnant-CP realization exhibits a stronger dependence on the
allowed Casas--Ibarra branch and the associated high-energy Yukawa
structure. Texture--branch combinations that are inefficient in one
RHN mass regime can become viable in another, while other combinations
do not reproduce the observed asymmetry within the corresponding
scanned parameter space.

The MMFGY framework therefore provides an economical setting in which low-energy neutrino observables can be connected with leptogenesis and characteristic correlations among the low-energy CP phases can be predicted, without imposing the full two-zero FGY texture. Improved measurements
of neutrino oscillation parameters and leptonic CP violation, together
with the increasing precision of reactor-neutrino experiments such as
JUNO, will further restrict the allowed low-energy parameter space and
sharpen the corresponding leptogenesis predictions. More generally, our results illustrate how minimal modifications of the Yukawa-sector constraints can preserve predictive relations between low- and high-energy neutrino physics, while leptogenesis imposes distinct constraints on the two realizations in different RHN mass regimes.

\section*{Acknowledgments}

This work was supported in part by the National Natural Science
Foundation of China under Grant Nos.~12475112 and 12447101,
Liaoning Revitalization Talents Program under Grant
No.~XLYC2403152, and the Basic Research Business Fees for
Universities in Liaoning Province under Grant
No.~LJ212410165050.

\bibliographystyle{unsrt}
\bibliography{references}

\end{document}